\documentclass[usenatbib]{mn2e}
\usepackage[]{graphicx}
\usepackage{amsmath,amssymb}
\usepackage{times}
\usepackage{subfigure}
\usepackage{epsfig}
\usepackage{graphics}

\renewcommand{\descriptionlabel}[1]%
  {\hspace{\labelsep}\textbf{#1}}

\title[Variable stars in NGC 6401]
     {Variable stars in the bulge globular cluster NGC~6401}

\author[Y. Tsapras et al.]
{\parbox[t]{\textwidth}{
Y. Tsapras$^{1,7}$, A. Arellano Ferro$^{2}$ ,D.M. Bramich$^{3}$, R. Figuera Jaimes$^{4,5}$, N. Kains$^{6}$,\\
R. Street$^{7}$, M. Hundertmark$^{8,5}$, K. Horne$^{5}$, M. Dominik$^{5}$, C. Snodgrass$^{9,10}$}
\\
\large{\it{(Affiliations can be found after the references)}}
}

\begin{document} 

\date{Draft June 2016}

\pagerange{\pageref{1}--\pageref{15}} \pubyear{2016}

\maketitle 

\label{firstpage}

\begin{abstract}
We present a study of variable stars in globular cluster NGC 6401. The cluster is only $5.3\degr$ away from the Galactic centre and suffers from strong differential reddening.
The photometric precision afforded us by difference image analysis resulted in improved sensitivity to variability in formerly inaccessible interior regions of the cluster. 
We find 23 RRab and 11 RRc stars within one cluster radius (2.4$\arcmin$), for which we provide coordinates, finder-charts and time-series photometry. 
Through Fourier decomposition of the RR Lyrae star light curves we derive a mean metallicity of [Fe/H]$_{\mathrm{UVES}} = -1.13 \pm 0.06$ (${\rm [Fe/H]}_{\mathrm{ZW}} = -1.25 \pm 0.06$), and a distance of $d\approx 6.35 \pm 0.81$ kpc. Using the RR Lyrae population, we also determine that NGC 6401 is an Oosterhoff type I cluster.
\end{abstract}
      
\begin{keywords}
Globular Clusters: NGC 6401 -- Variable Stars: RR Lyrae.
\end{keywords}

\section{Introduction}
Globular clusters are gravitationally bound stellar systems which, for the most part, contain between $10^5$ to $10^6$ stars in a volume spanning a few hundred parsecs. These systems are home to some of the oldest stellar populations in the Galaxy and are rich in variable stars of the RR Lyrae type, which are particularly useful for obtaining distance measurements. There are currently 158 globular clusters associated with the Milky Way for which there exist estimates of their distance, metallicity, age and dynamical parameters \citep{Harris96}.

This paper belongs to a series of studies of variable stars in globular clusters (e.g. \citet{Ferro15,Kains15,Ferro13,Figuera13,Bramich11}) aimed primarily at Fourier decomposing the light curves of RR Lyrae stars in search of physical parameters and at updating the variability census using modern instruments and photometric reduction methods. In particular, the combination of modern CCDs and the technique of difference image analysis \citep{Alard98,Bramich08} allow us to probe for variability closer to the crowded central regions of the clusters than was previously possible.

Globular cluster NGC~6401 (also known as GCL1735-238 and ESO520-SC11) is located at equatorial coordinates $(\alpha,\delta)_{J2000} = (17^{\mbox{\scriptsize h}}38^{\mbox{\scriptsize m}}37^{\mbox{\scriptsize s}}, -23\degr 54\arcmin 32\arcsec )$ and galactic coordinates $(l,b) = (3.45\degr, +3.98\degr)$, only $5.3\degr$ from the Galactic centre. According to the 2010 update to the \citet{Harris96} catalogue, it is a relatively metal rich cluster ([Fe/H]$\approx -1.02$ dex) with an estimated distance from the Sun of $d_{\odot}\approx 10.6$kpc. However, the metallicity estimate is contradicted by studies done by \citet{Davidge01} and \citet{Valenti07}, who derive metallicities of [Fe/H]$\approx -1.5$ and [Fe/H]$\approx -1.37$ dex respectively. They conclude that the cluster appears to be metal poor.

The paper is structured as follows: In section \ref{sec:Observations}, we give an account of our observations and data reduction methods. Section \ref{sec:variables} contains a description of our methodological approach to identifying and classifying variable sources in the vicinity of NGC~6401. In section \ref{sec:Physical} we provide an account of the Fourier decomposition of the light curves of the RR Lyrae stars and how the derived parameters are used to estimate the physical properties of the cluster. We comment on our results in section \ref{sec:discussion} and conclude in section \ref{sec:summary}.

\section{Observations and Reductions}
\label{sec:Observations}

\subsection{Observations}
Observations of the target field were a by-product of the RoboNet microlensing campaign \citep{Tsapras09} and were obtained in the campaign-specific SDSS-$g\arcmin$ and SDSS-$i\arcmin$ bands throughout the months of July and August 2013 using the LCOGT\footnote{www.lcogt.net} 1m robotic telescope network. The LCOGT telescopes are clones of each other, featuring identical instruments, and are deployed in clusters of two to three at the Siding Spring Observatory (SSO), South African Astronomical Observatory (SAAO) and Cerro Tololo Inter-American Observatory (CTIO). The detectors used were standard SBIG STX-16803, with 4096$\times$4096 pixels, each 0.23$\arcsec$, giving a field of view of $15.8\arcmin \times 15.8\arcmin$. Image binning was set to 2$\times$2, giving an effective pixel scale of 0.47$\arcsec$. A summary of the observations can be found in table~\ref{tab:obs}. 

\begin{table}
\scriptsize
\caption{The distribution of observations of NGC 6401 for each filter, where the
columns $N_{g}$ and $N_{i}$ represent the number of images taken in the  SDSS-$g\arcmin$ and SDSS-$i\arcmin$
bands respectively. We also provide the exposure time(s)
employed during each night for each filter in the columns $t_{g}$ and $t_{i}$.}
\centering
\begin{tabular}{lccccc}
\hline
Date & $N_{g}$ & $t_{g}$ (s) & $N_{i}$ & $t_{i}$ (s) \\
\hline
20130705 & --- & ---	& 2  & 200 \\
20130707 & --- & ---	& 8  & 200 \\
20130708 & --- & ---	& 7  & 200 \\
20130711 & --- & ---	& 4  & 200 \\
20130712 & --- & ---	& 1  & 200 \\
20130717 & --- & ---	& 4  & 200 \\
20130721 & 41  & 60-200 & 48 & 60-200 \\
20130722 & 8   & 60-200 & 4  & 60 \\
20130724 & 21  & 60-200 & 29 & 60-200 \\
20130725 & 39  & 60-200 & 41 & 60-200 \\
20130726 & 34  & 60-200 & 22 & 60-200 \\
20130727 & 40  & 60-200 & 43 & 60-200 \\
20130728 & 13  & 60-200 & 39 & 60-200 \\
20130729 & 22  & 60-200 & 39 & 60-200 \\
20130730 & 19  & 60-200 & 18 & 60-200 \\
20130731 & 25  & 60-200 & 27 & 60-200 \\
20130801 & 27  & 60-200 & 32 & 60-200 \\
20130802 & 9   & 60-200 & 12 & 60-200 \\
20130803 & 22  & 60-200 & 10 & 60-200 \\
20130804 & 22  & 60-200 & 18 & 60-200 \\
20130805 & 14  & 60-200 & 20 & 60-200 \\
20130808 & 2   & 60	& 4  & 60-200 \\
20130809 & 39  & 60-200 & 36 & 60-200 \\
20130810 & 16  & 60-200 & 23 & 60-200 \\
20130811 & 12  & 60-200 & 18 & 60-200 \\
\hline
Total:   & 425  &	& 509 &       \\
\hline
\end{tabular}
\label{tab:obs}
\end{table}

\subsection{Difference Image Analysis}
Using our {\tt DanDIA}\footnote{{\tt DanDIA} is built from the DanIDL library of IDL routines available at http://www.danidl.co.uk} pipeline \citep{Bramich13}, we performed difference image analysis (DIA) on the images to extract high-precision photometric measurements. Since all images were obtained from a network of identical telescopes, all images taken with the same filter were processed together, regardless of which telescope they originated from.

At the first step, every image underwent standard debiasing and flat-fielding calibrations. Subsequently, a reference (template) image was constructed for each filter by combining a set of registered best-seeing images obtained with the same instrument. For each reference image we measured the fluxes and positions of all PSF-like objects (stars) by extracting a spatially variable empirical PSF from the image and fitting this PSF to each detected object. Image registration was performed using the triangle matching method of \citet{Pal06} to match the stars detected in each image with those detected in the reference, and then deriving a linear transformation between the images that uses cubic O-MOMS resampling \citep{Blu01}. In this way, all images were geometrically aligned to match the relevant reference. 

Difference images were generated by subtracting the relevant reference image, convolved with an appropriate spatially variable kernel, from each registered image. The spatially variable convolution kernel was determined using bilinear interpolation of a set of kernels, modelled as discrete pixel arrays \citep{Bramich08}, that were derived for a uniform 13$\times$13 grid of subregions across the image.

Difference fluxes for each star detected in the reference image were evaluated from each difference image by using fixed position PSF photometry. Light curves were constructed by calculating the total flux $f_{\mbox{\scriptsize tot}}(t)$ in ADU/s at each epoch $t$ using:
\begin{equation}
f_{\mbox{\scriptsize tot}}(t) = f_{\mbox{\scriptsize ref}} +
\frac{f_{\mbox{\scriptsize diff}}(t)}{p(t)}
\label{eqn:totflux}
\end{equation}
where $f_{\mbox{\scriptsize ref}}$ is the reference flux (ADU/s),
$f_{\mbox{\scriptsize diff}}(t)$ is the differential flux (ADU/s) and
$p(t)$ is the photometric scale factor (the integral of the kernel solution).

To convert to instrumental magnitudes, we used:
\begin{equation}
m_{\mbox{\scriptsize ins}}(t) = 25.0 - 2.5 \log \left[ f_{\mbox{\scriptsize tot}}(t) \right]
\label{eqn:mag}
\end{equation}
where $m_{\mbox{\scriptsize ins}}(t)$ is the instrumental magnitude of the star 
at time $t$. Uncertainties were propagated in the appropriate analytical fashion.
\begin{figure} 
\includegraphics[width=9.0cm]{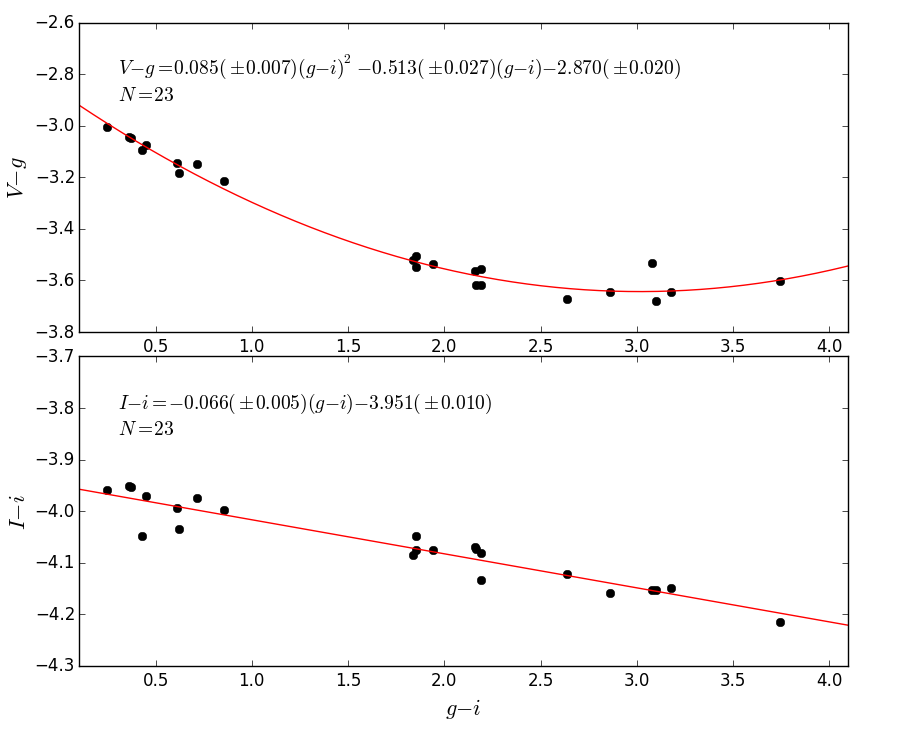}
\caption{Transformation relations between the instrumental and the standard
photometric systems using a set of standard stars in the field of NGC~6401 provided
by Peter Stetson.}
\label{transV}
\end{figure}

The above procedure and its caveats have been described in detail in \citet{Bramich11} and the interested reader is referred there for relevant details.

\subsection{Photometric Calibrations}

\subsubsection{Relative}
\label{sec:rel}
Systematic errors in photometric measurements are caused by a whole range of effects such as mismatches between the true and measured flat-field calibrations, insufficiently accurate modelling of the spatially varying PSF, poor image subtractions in DIA, etc. These types of errors affect all light curves to a greater or lesser extent dependent on the data properties and photometry algorithms employed. Although it is not possible to suppress them completely, it is often the case that substantial qualitative improvements can be made when dealing with time-series photometry (e.g. \citet{Bramich15,Kains15}). 
We used the method from \citet{Bramich12_2} to determine the magnitude offsets $Z_{k}$ that should be applied to each photometric measurement from the image $k$, which translates into a correction for any systematic errors introduced into the DIA photometry due to errors in the fitted photometric scale factors $p(t)$. The offsets derived were of order $\sim$1-2\% for the SDSS-$g\arcmin$ band, and somewhat smaller for the SDSS-$i\arcmin$ band. The application of this correction substantially improved the quality of the light curves.

\subsubsection{Absolute}
\label{absolute}

We identified 23 standard stars in the field of NGC~6401 from the online collection of
Stetson (2000)\footnote{www3.cadc-ccda.hia-iha.nrc-cnrc.gc.ca/community/STETSON/standards}
which we used to transform instrumental ($g, i$) magnitudes into the standard
($V, I$) system. Our derived transformations are depicted in Figure~\ref{transV}. For the $i$-band we obtained a linear relation, as expected, since the SDSS-$i\arcmin$ band is similar to the Johnson-Cousins $I$-band. However, for the observations in the $g$-band, we obtained a non-linear (quadratic) relation.  
This is due to the fact that the SDSS-$g\arcmin$ band is bluer than the Johnson-Cousins $V$-band \citep{Ivezic07}. Furthermore, NCG~6401 is heavily affected by differential reddening which is more evident in the bluer bands. 
The analytical transformations used for the calibration are as follows:
\begin{equation}
\begin{aligned}
V_{std} = g + 0.085(\pm 0.007)(g-i)^2 - 0.513(\pm 0.027)(g-i) \\
          - 2.870(\pm 0.020),
\end{aligned}
\end{equation}
\begin{equation}
I_{std} = i - 0.066(\pm 0.005)(g-i) - 3.951(\pm 0.010).
\label{eqn:istd}
\end{equation}
It is important to note that while these transformations shift the instrumental magnitudes to the $V$ and $I$ magnitude scales, the fact that the photometry pertains to the $g$ and $i$-bands remains unchanged. Hence the transformations do not affect the overall shape/amplitude of the light curves for the variable stars in our sample, and the light curves we present are simply $g$ and $i$-band light curves on the standard $V$ and $I$ brightness scale.
 
\begin{figure}
 \centering
 \includegraphics[width=8.5cm]{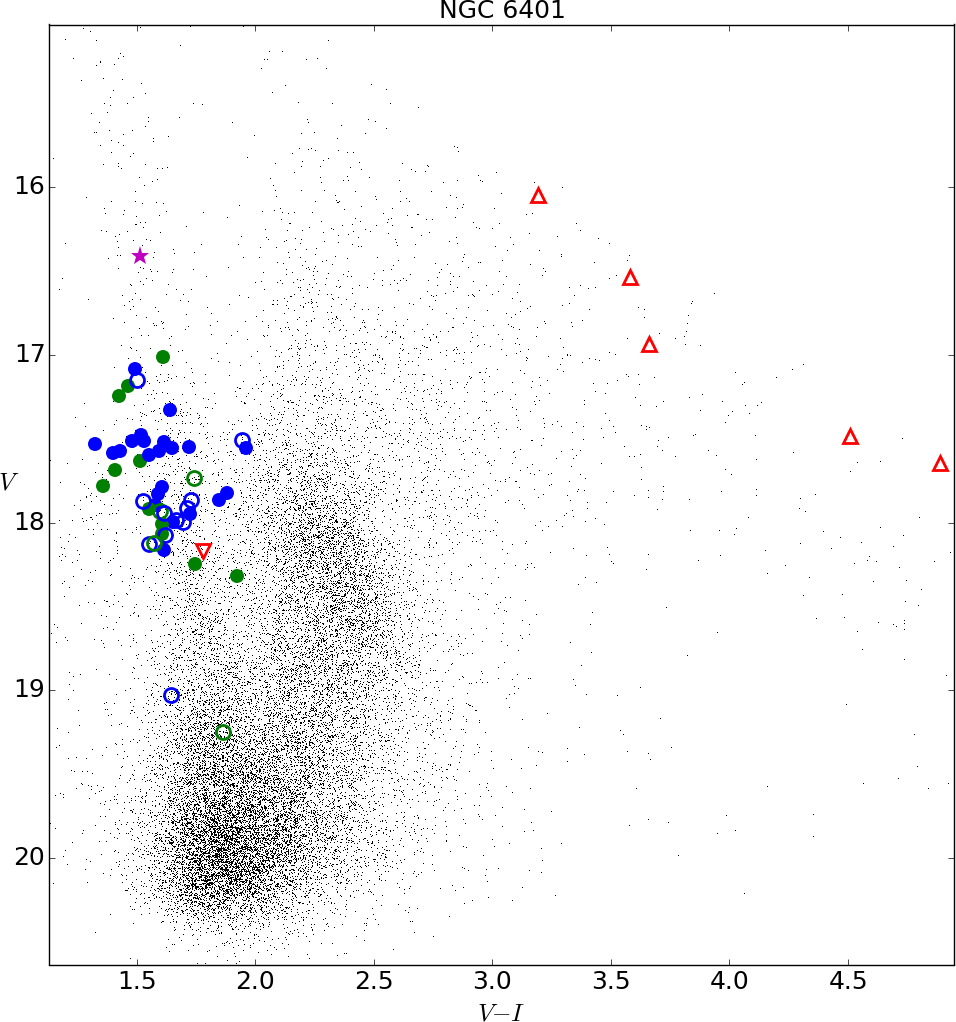}
 \caption{Colour-magnitude diagram of NGC 6401. Blue and green filled circles mark the cluster RRab and RRc stars respectively. Unfilled circles represent non-cluster RR Lyraes. The triangles on the top right are long period variables NSV09307, NSV09288, LPV9, LPV10 and LPV11. The purple star marks the location of V3. The inverted red triangle on the middle-left of the plot ($V \sim 18.1$) is variable E1.}
 \label{cmd}
\end{figure}

Once these transformations were applied to the light curves, we proceeded to generate the colour-magnitude diagram (CMD) shown in Figure~\ref{cmd} using the inverse-variance-weighted mean $V$ and $I$ magnitudes for 23698 stars identified in our images. Blue and green filled circles mark the relevant locations on the diagram for the cluster RRab and RRc stars respectively (see section~\ref{sec:variables}). The plot shows an almost vertical main sequence (MS) between magnitudes 15 to 20, which is mostly due to contamination from foreground stars in the Galactic disk. The red horizontal branch (RHB) is the concentration of stars at $V\sim 18$, which is mixed with the fainter end of the red giant branch (RGB). The tilted RGB extends to magnitude $V-I\sim 5-6$. Some differential reddening is present, indicated by the tilt and elongation of the RHB and RGB. 

To validate the transformations, we compare our calibrated magnitudes with a set of 1041 photometrically calibrated OGLE stars from the field BLG626.24 in the vicinity of NGC~6401\footnote{The OGLE light curves were kindly provided by Igor Soszynski.} and find that for the majority of these stars the magnitudes match to within $\sim$0.06 mag.

\subsection{Astrometry}
\label{sec:astrometry}

We used the {\tt GAIA} \citep{Draper00} image display tools to derive a linear astrometric solution for the SDSS-$g\arcmin$ filter reference image by matching 1429 stars from the 
UCAC4 star catalogue \citep{Zacharias12}. After manually rejecting saturated stars and stars at the edges of the image we achieved a radial RMS scatter in the residuals of
$\sim$0.22$\arcsec$. Similarly, for the SDSS-$i\arcmin$ filter we used 1528 stars from the UCAC4 catalogue and obtained an RMS of $\sim$0.25$\arcsec$. The astrometric fit was then used to calculate the J2000.0 celestial coordinates for all of the confirmed variables in our field of view (see Tables~\ref{tab:var} and \ref{tab:var_out}). 

\section{Variable stars in NGC~6401}
\label{sec:variables}
A series of searches for variable stars at and around NGC~6401 conducted in the '70s by \citet{Terzan72,Terzan73,Terzan75} yielded 25 variables (labelled V1-25) that could be cluster members\footnote{They also identified a large number of variable star candidates in our field of view that are not cluster members, but we have not been able to recover any of them in our search.}. They are listed in the \citet{Clement01} catalogue in the January 2015 update\footnote{http://www.astro.utoronto.ca/~cclement/cat/C1735m238}. However, detailed studies of these variables were never published.
\begin{table*}
\caption{Time-series SDSS-$g^{\prime}$ and SDSS-$i^{\prime}$ photometry, linearly transformed to the V and I bands respectively, for all of the confirmed variables in our field of view. The standard $M_{\mbox{\scriptsize std}}$ and instrumental $m_{\mbox{\scriptsize ins}}$ magnitudes are listed in columns 4 and 5, respectively, corresponding to the variable star, filter, and epoch of mid-exposure listed in columns 1-3, respectively. The uncertainty on $m_{\mbox{\scriptsize ins}}$ is listed in column 6, which also corresponds to the uncertainty on $M_{\mbox{\scriptsize std}}$. For completeness, we also list the quantities $f_{\mbox{\scriptsize ref}}$, $f_{\mbox{\scriptsize diff}}$ and $p$ from Equation~\ref{eqn:totflux} in columns 7, 9 and 11, along with the uncertainties $\sigma_{\mbox{\scriptsize ref}}$ and $\sigma_{\mbox{\scriptsize diff}}$ in columns 8 and 10. This is an extract from the full table, which is available with the electronic version of the article.}
\centering
\begin{tabular}{ccccccccccc}
\hline
Variable & Filter & HJD & $M_{\mbox{\scriptsize std}}$ & $m_{\mbox{\scriptsize ins}}$ & $\sigma_{m}$ & $f_{\mbox{\scriptsize ref}}$ & $\sigma_{\mbox{\scriptsize ref}}$ & $f_{\mbox{\scriptsize diff}}$ & $\sigma_{\mbox{\scriptsize diff}}$ & $p$ \\
Star ID  &        & (d) & (mag)                        & (mag)                        & (mag)        & (ADU s$^{-1}$)         & (ADU s$^{-1}$)  & (ADU s$^{-1}$) & (ADU s$^{-1}$) &    \\
\hline
V2 & $g$ & 2456495.20657  & 18.024 & 21.286 & 0.145 & 34.473 & 0.924 & -6.838 & 7.218 & 1.7640 \\
V2 & $g$ & 2456495.21746  & 17.956 & 21.218 & 0.110 & 34.473 & 0.924 & -3.424 & 5.963 & 1.8019 \\
\vdots   & \vdots & \vdots  & \vdots & \vdots & \vdots & \vdots   & \vdots & \vdots   & \vdots & \vdots \\
V2 & $i$ & 2456479.64343 &  16.088 &  20.099 & 0.011 & 103.089 & 2.051 & -24.881 & 1.968 & 2.1117 \\
V2 & $i$ & 2456479.74302 &  16.186 &  20.196 & 0.020 & 103.089 & 2.051 & -12.404 & 0.994 & 0.6315 \\
\vdots   & \vdots & \vdots  & \vdots & \vdots & \vdots & \vdots   & \vdots & \vdots   & \vdots & \vdots \\
\hline
\end{tabular}
\label{tab:gi_phot}
\end{table*}

\begin{table*}
\scriptsize
\caption{Details for all confirmed RR Lyrae stars within one cluster radius (2.4$\arcmin$) from the NGC~6401 cluster centre. Period estimates for each variable from this work are reported in column 4 and from previous studies in column 5 for comparison. Mean magnitudes are reported in columns 8 and 9 and peak-to-peak amplitudes in columns 10 and 11. The epoch reported in column 12 is the light curve maximum.}
\label{tab:var}
\centering
\begin{tabular}{llllllllllll}
\hline
 			 &	   &		    &	      & OGLE	  &		&	      & 	    &		  &		&	      & 	      \\
Variable		 & OGLE ID & Variable	    & Period  & Period    &  RA 	&  Dec        & $<{\it V}>$ & $<{\it I}>$ & A$_{g}$	& A$_{i}$     & Epoch	      \\
Star ID 		 &	   & Type	    & (days)  & (days)    & (J2000.0)	& (J2000.0)   & (mag)	    & (mag)	  & (mag)	& (mag)       & (HJD-2450000) \\
\hline
V2  & OGLE-BLG-RRLYR-24004 & RRab	    & 0.70158 &  0.70235  & 17:38:33.97 & -23:54:26.6 & 17.66       &  15.94	   & 0.93	 & 0.48       & 6500.9323     \\
V4  & OGLE-BLG-RRLYR-24049 & RRab	    & 0.47233 &  0.47248  & 17:38:36.36 & -23:54:35.2 & ---	    &  ---         & 1.18	 & ---        & 6499.2669     \\
V5  & OGLE-BLG-RRLYR-24006 & RRab	    & 0.53289 &  0.53386  & 17:38:34.01 & -23:54:40.8 & 18.23       &  16.39	   & 1.19	 & 0.64       & 6514.9621     \\
V6  & OGLE-BLG-RRLYR-24078 & RRab	    & 0.44446 &  0.44417  & 17:38:38.19 & -23:53:58.2 & 17.96       &  16.37	   & 1.55	 & 0.91       & 6507.3248     \\
V7  & OGLE-BLG-RRLYR-24126 & RRab	    & 0.54563 &  0.54575  & 17:38:40.99 & -23:54:32.3 & 17.60       &  16.00	   & 1.19	 & 0.62       & 6495.2935     \\
V8  & ---		   & RRc	    & 0.32485 &  ---	  & 17:38:38.72 & -23:55:25.0 & 17.71       &  16.13	   & 0.54	 & 0.24       & 6508.3582     \\
V9  & OGLE-BLG-RRLYR-24055 & RRab	    & 0.51367 &  0.51311  & 17:38:36.76 & -23:54:39.2 & 17.51       &  15.75	   & 1.15	 & 0.54       & 6505.9660     \\
V10 & ---		   & RRab	    & 0.52603 &  ---	  & 17:38:38.57 & -23:53:52.4 & 17.88       &  16.20	   & 1.32	 & 0.68       & 6503.5181     \\
V11 & OGLE-BLG-RRLYR-24121 & RRab	    & 0.58212 &  0.58183  & 17:38:40.82 & -23:54:08.2 & 17.59	    &  16.00	   & 0.90	 & 0.46       & 6501.5939     \\
V12 & OGLE-BLG-RRLYR-24030 & RRc	    & 0.29130 &  0.29109  & 17:38:35.09 & -23:54:30.8 & 17.23	    &  15.73	   & 0.67	 & 0.34       & 6503.5043     \\
V13 & OGLE-BLG-RRLYR-24036 & RRab	    & 0.56498 &  0.56527  & 17:38:35.63 & -23:54:49.0 & 18.04	    &  16.31	   & 1.13	 & 0.59       & 6495.4983     \\
V14 & ---	           & RRab	    & 0.60851 &  ---	  & 17:38:39.37 & -23:54:39.0 & 17.58	    &  15.84	   & 0.54	 & 0.24       & 6495.3499     \\
V15 & OGLE-BLG-RRLYR-24075 & RRab	    & 0.50440 &  0.50644  & 17:38:37.78 & -23:55:09.5 & 17.84	    &  16.21	   & 1.30	 & 0.71       & 6499.2100     \\
V16 & OGLE-BLG-RRLYR-24043 & RRab	    & 0.50581 &  0.50552  & 17:38:36.00 & -23:55:58.1 & 18.51	    &  16.63	   & 1.27	 & 0.70       & 6505.5065     \\
V17 & OGLE-BLG-RRLYR-24131 & RRab	    & 0.50517 &  0.50524  & 17:38:41.67 & -23:52:57.0 & 17.22	    &  15.62	   & 1.34	 & 0.74       & 6501.7732     \\
V18 & OGLE-BLG-RRLYR-24002 & RRab	    & 0.56830 &  0.57021  & 17:38:33.84 & -23:54:03.8 & 17.71	    &  16.05	   & 1.25	 & 0.66       & 6499.5121     \\
V19 & OGLE-BLG-RRLYR-24160 & RRab	    & 0.55989 &  0.55949  & 17:38:43.80 & -23:56:12.1 & 17.63	    &  16.04	   & 1.06	 & 0.57       & 6507.4949     \\
V21 & OGLE-BLG-RRLYR-24056 & RRc	    & 0.27809 &  0.27814  & 17:38:36.77 & -23:54:12.2 & 17.86	    &  16.46	   & 0.52	 & 0.30       & 6507.9500     \\
V22 & OGLE-BLG-RRLYR-24025 & RRc	    & 0.29972 &  0.29971  & 17:38:34.87 & -23:54:40.6 & 17.98	    &  16.37	   & 0.50	 & 0.24       & 6499.2900     \\
V23 & OGLE-BLG-RRLYR-24100 & RRc	    & 0.36641 &  0.36647  & 17:38:39.91 & -23:52:27.2 & 17.28	    &  15.82	   & 0.51	 & 0.25       & 6501.0665     \\
V24 & OGLE-BLG-RRLYR-24182 & RRab	    & 0.52000 &  0.52013  & 17:38:45.53 & -23:53:54.8 & 17.65	    &  16.01	   & 1.05	 & 0.55       & 6507.9216     \\
V25 & OGLE-BLG-RRLYR-24047 & RRab	    & 0.55867 &  0.55860  & 17:38:36.27 & -23:55:03.0 & 17.55	    &  15.59	   & 0.71	 & 0.28       & 6504.6719     \\
V26 & OGLE-BLG-RRLYR-23926 & RRab	    & 0.49585 &  0.49683  & 17:38:28.16 & -23:54:24.2 & 18.22	    &  16.33	   & 1.16	 & 0.59       & 6509.9683     \\
V27 & OGLE-BLG-RRLYR-23985 & RRc	    & 0.28743 &  0.28783  & 17:38:33.01 & -23:55:04.8 & 18.40	    &  16.40	   & 0.55	 & 0.19       & 6503.4900     \\
V28 & OGLE-BLG-RRLYR-23999 & RRab	    & 0.62576 &  0.62373  & 17:38:33.56 & -23:56:52.8 & 18.00	    &  16.24	   & 0.54	 & 0.34       & 6499.5121     \\
V29 & OGLE-BLG-RRLYR-24011 & RRc	    & 0.28860 &  0.28856  & 17:38:34.16 & -23:55:07.2 & 18.34	    &  16.52	   & 0.55	 & 0.24       & 6503.0370     \\
V30 & OGLE-BLG-RRLYR-24021 & RRc	    & 0.26198 &  0.26207  & 17:38:34.79 & -23:55:05.0 & 18.07	    &  16.42	   & 0.42	 & 0.20       & 6498.4785     \\
V31 & OGLE-BLG-RRLYR-24024 & RRab	    & 0.57712 &  0.57840  & 17:38:34.87 & -23:54:35.6 & 18.00	    &  16.06	   & 0.91	 & 0.39       & 6495.4983     \\
V32 & OGLE-BLG-RRLYR-24037 & RRab	    & 0.66136 &  0.66059  & 17:38:35.65 & -23:54:51.8 & 17.94	    &  15.96	   & 0.68	 & 0.30       & 6495.2728     \\
V33 & OGLE-BLG-RRLYR-24039 & RRc	    & 0.34032 &  0.34087  & 17:38:35.70 & -23:55:26.3 & 18.13	    &  16.48	   & 0.50	 & 0.25       & 6503.6545     \\
V34 & OGLE-BLG-RRLYR-24060 & RRab	    & 0.54441 &  0.54533  & 17:38:36.79 & -23:54:35.4 & 17.60	    &  15.93	   & 0.90	 & 0.41       & 6505.6228     \\
V35 & OGLE-BLG-RRLYR-24062 & RRc	    & 0.28284 &  0.28300  & 17:38:37.00 & -23:54:35.5 & 17.05	    &  15.41	   & 0.31	 & 0.12       & 6501.6900     \\
V36 & OGLE-BLG-RRLYR-24142 & RRc	    & 0.31898 &  0.31905  & 17:38:42.86 & -23:52:57.3 & 17.76	    &  16.29	   & 0.44	 & 0.23       & 6500.5692     \\
V37 & OGLE-BLG-RRLYR-24148 & RRab	    & 0.47915 &  0.47961  & 17:38:43.06 & -23:52:35.3 & 17.93	    &  16.23	   & 1.38	 & 0.77       & 6500.0600     \\
\hline
\end{tabular}
\end{table*}

\begin{table*}
\scriptsize
\caption{Details for all remaining confirmed variables in the field of NGC~6401. Period estimates for each variable from this work are reported in column 3 and from previous studies in column 4 for comparison. Mean magnitudes are reported in columns 7 and 8. The epoch reported in column 9 is the light curve maximum.}
\label{tab:var_out}
\centering
\begin{tabular}{lllllllll}
\hline
         &          	  &	   & OGLE   &		 &	     & & & \\
Variable & Variable 	  & Period & Period &  RA	 & Dec       & $<{\it V}>$& $<{\it I}>$ & Epoch \\
Star ID  & Type    	  & (days) & (days) &  (J2000.0) & (J2000.0) & (mag) & (mag) & (HJD-2450000) \\
\hline
V3		    	 & CWB      	      & 1.74869   &  ---      & 17:38:41.31 & -23:54:34.9 & 16.41  &  14.90 & 6497.9440 \\
E1		    	 & EB   	      & 1.33228   &  ---      & 17:38:37.42 & -24:00:52.0 & 18.17  &  16.39 & 6495.4922 \\
LPV1		    	 & LPV		      & ---	  &  ---      & 17:38:41.02 & -23:56:26.6 & ---    &  ---   & 6495.4983 \\
LPV2		    	 & LPV  	      & ---	  &  ---      & 17:38:36.89 & -23:49:23.5 & ---    &  ---   & 6509.6726 \\
LPV3		    	 & LPV  	      & ---	  &  ---      & 17:38:36.35 & -23:54:20.5 & ---    &  ---   & 6509.2758 \\
LPV4		    	 & LPV  	      & ---	  &  ---      & 17:38:33.28 & -23:58:37.9 & ---    &  ---   & 6498.5154 \\
LPV5		    	 & LPV  	      & ---	  &  ---      & 17:38:31.02 & -23:48:47.8 & ---    &  ---   & 6501.7620 \\
LPV6		    	 & LPV  	      & ---	  &  ---      & 17:38:29.84 & -23:57:17.1 & ---    &  ---   & 6503.5181 \\
LPV7		    	 & LPV  	      & ---	  &  ---      & 17:38:23.66 & -23:51:26.8 & ---    &  ---   & 6507.5209 \\
LPV8		    	 & LPV  	      & ---	  &  ---      & 17:38:18.85 & -23:51:04.2 & ---    &  ---   & 6501.7620 \\
LPV9		    	 & LPV  	      & ---	  &  ---      & 17:38:15.93 & -24:00:04.6 & 16.54  &  12.96 & 6514.6812 \\
LPV10		    	 & LPV  	      & ---	  &  ---      & 17:38:12.37 & -23:51:55.4 & 16.93  &  13.27 & 6516.5155 \\
LPV11		    	 & LPV  	      & ---	  &  ---      & 17:38:08.05 & -23:47:16.6 & 16.04  &  12.85 & 6514.4772 \\
LPV12		    	 & LPV  	      & ---	  &  ---      & 17:38:41.74 & -23:54:22.4 & ---    &  ---   & 6495.2174 \\
NSV 09288		 & LPV  	      & ---	  &  ---      & 17:38:02.73 & -23:46:41.4 & 17.48  &  12.97 & 6514.4762 \\
NSV 09307		 & LPV  	      & ---	  &  ---      & 17:38:17.70 & -24:01:29.9 & 17.65  &  12.75 & 6495.5296 \\
NSV 09349		 & LPV		      & ---	  &  ---      & 17:38:46.09 & -23:47:33.3 & ---    &  ---   & 6503.5034 \\
V2226 Oph		 & LPV  	      & ---	  &  ---      & 17:38:21.23 & -23:50:43.7 & ---    &  ---   & 6503.5181 \\
V2227 Oph		 & LPV  	      & ---	  &  ---      & 17:38:21.80 & -23:55:22.7 & ---    &  ---   & 6516.5146 \\
OGLE-BLG-RRLYR-23731	 & RRab 	      & 0.68325   &  0.68279  & 17:38:09.55 & -24:01:35.9 & 17.87  &  16.34 & 6495.2390 \\
OGLE-BLG-RRLYR-23799	 & RRab 	      & 0.52350   &  0.52347  & 17:38:16.60 & -23:46:36.7 & ---    &  ---   & 6495.3700 \\
OGLE-BLG-RRLYR-23863	 & RRab?  	      & 0.68212   &  0.68160  & 17:38:22.97 & -23:51:51.6 & 17.74  &  16.00 & 6498.5800 \\
OGLE-BLG-RRLYR-23880	 & RRab 	      & 0.55167   &  0.55149  & 17:38:24.43 & -23:51:49.2 & 17.98  &  16.32 & 6495.7350 \\
OGLE-BLG-RRLYR-23969	 & RRc  	      & 0.32692   &  0.32682  & 17:38:31.63 & -23:50:06.4 & 19.27  &  17.40 & 6495.8150 \\
OGLE-BLG-RRLYR-24007	 & RRab 	      & 0.44022   &  0.43985  & 17:38:34.08 & -23:59:28.3 & 19.03  &  17.38 & 6497.4737 \\
OGLE-BLG-RRLYR-24026	 & RRab 	      & 0.50640   &  0.50689  & 17:38:34.89 & -23:57:13.5 & 17.99  &  16.33 & 6495.1150 \\
OGLE-BLG-RRLYR-24044	 & RRab               & 0.46785   &  0.46743  & 17:38:36.09 & -23:50:17.5 & 18.02  &  16.30 & 6509.9880 \\
OGLE-BLG-RRLYR-24050	 & RRc  	      & 0.36286   &  0.36284  & 17:38:36.46 & -24:01:15.3 & 18.12  &  16.55 & 6501.9897 \\
OGLE-BLG-RRLYR-24058	 & RRab  	      & 0.60849   &  0.60920  & 17:38:36.79 & -23:51:51.3 & 17.87  &  16.14 & 6500.7392 \\
OGLE-BLG-RRLYR-24066	 & RRab 	      & 0.51589   &  0.51623  & 17:38:37.43 & -23:47:53.4 & ---    &  ---   & 6500.6292 \\
OGLE-BLG-RRLYR-24124     & RRab               & 0.49729   &  0.49800  & 17:38:40.86 & -23:51:59.3 & 17.15  &  15.66 & 6509.0082 \\
OGLE-BLG-RRLYR-24138	 & RRc  	      & 0.31636   &  0.31635  & 17:38:42.60 & -24:01:22.4 & 17.93  &  16.34 & 6507.7153 \\
OGLE-BLG-RRLYR-24255     & RRab  	      & 0.61989   &  0.61915  & 17:38:50.64 & -23:54:11.7 & 17.51  &  15.57 & 6496.2955 \\
OGLE-BLG-RRLYR-24293     & RRab 	      & 0.53190   &  0.53177  & 17:38:53.97 & -23:53:30.3 & 18.07  &  16.45 & 6496.2615 \\
OGLE-BLG-RRLYR-24325     & RRab 	      & 0.51808   &  0.51817  & 17:38:56.67 & -23:56:19.0 & 18.13  &  16.58 & 6499.2784 \\
OGLE-BLG-RRLYR-24383     & RRab 	      & 0.61407   &  0.61420  & 17:39:02.24 & -24:00:54.3 & 17.91  &  16.20 & 6504.9341 \\
\hline
\end{tabular}
\end{table*}
The first observations of this cluster using CCDs were performed by \citet{Barbuy99}, who observed for a single night (July 4, 1998) with the 1.5m Danish telescope at ESO (La Silla). Although theirs was not a search for variables, they performed PSF-fitting photometry with Daophot II \citep{Stetson87} on their images and used their CMD diagrams to estimate the amount of reddening and the cluster's distance from the Sun. 

NGC~6401's proximity to the Galactic bulge places it in the fields monitored by the fourth phase of the Optical Gravitational Lensing Experiment (OGLE) survey which started operations in 2011 \citep{Udalski15}. \citet{Soszynski14} present a comprehensive collection of $\sim$38000 RR Lyrae stars detected over 182 square degrees monitored by the OGLE-IV survey in the densest regions of the Galactic bulge. Of those, $\sim$300 are identified as plausible members of 15 globular clusters. Their list includes 31 RR Lyrae stars lying within one cluster angular radius from the centre of NGC~6401, several of which were previously unknown. Their photometric reductions were performed using the OGLE real time photometric pipeline \citep{Udalski03}, which is based on difference image analysis, and the final photometry of selected variables was calibrated to the standard $VI$ system with an estimated zero-point accuracy of $\sim$0.02mag.
\begin{figure*}
 \centering
 \includegraphics[width=17.0cm]{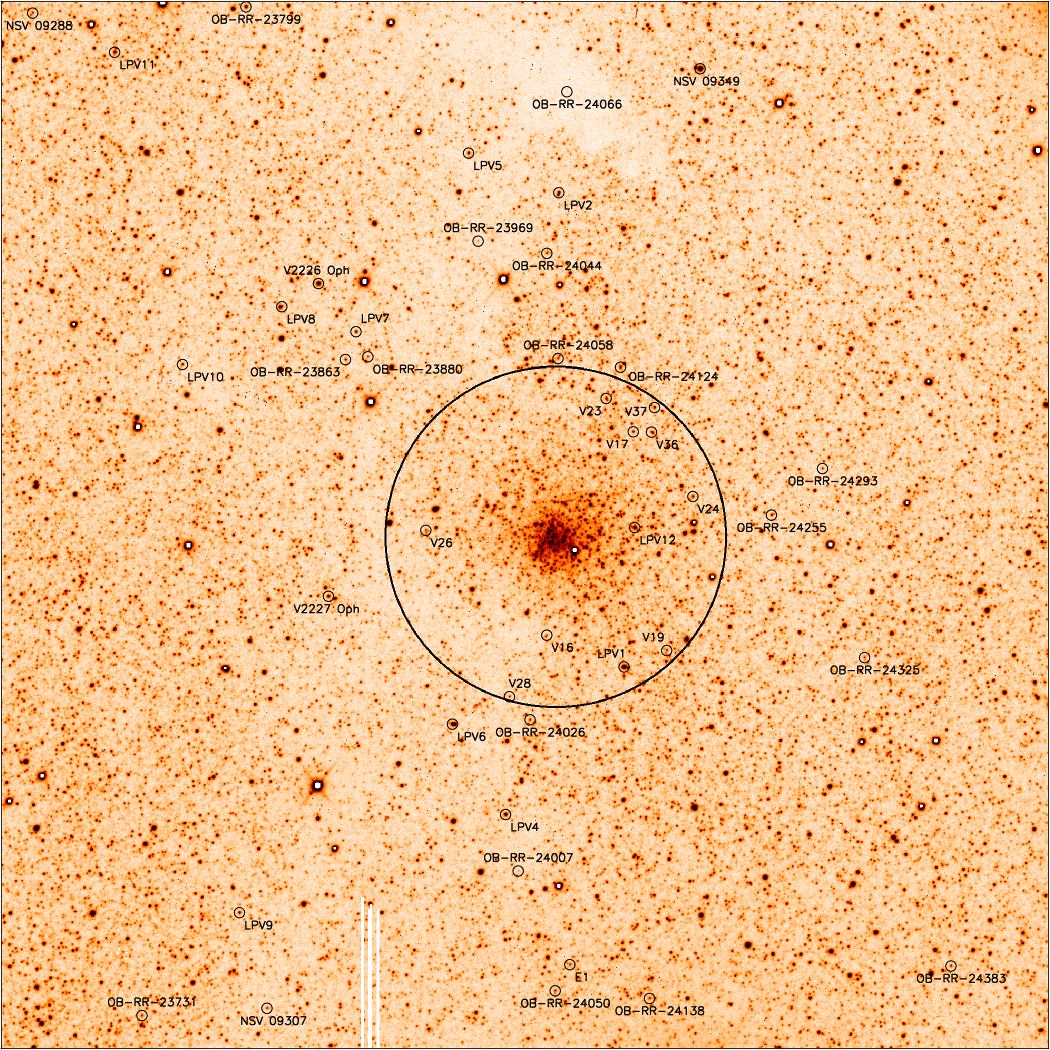}
 \caption{Finding chart constructed from our SDSS-$g\arcmin$ band reference image. North is up and East is to the right. The cluster image is 15.8$\times$15.8 arcmin$^{2}$. All of the confirmed variables in our FoV are identified, except in the crowded cluster core (see Figure~\ref{fchart_centre}). The circle marks the cluster radius of 2.4$\arcmin$. OGLE variable names have been abbreviated for clarity.}
 \label{fchart_full}
\end{figure*}

\begin{figure*}
 \centering
 \includegraphics[width=16.0cm]{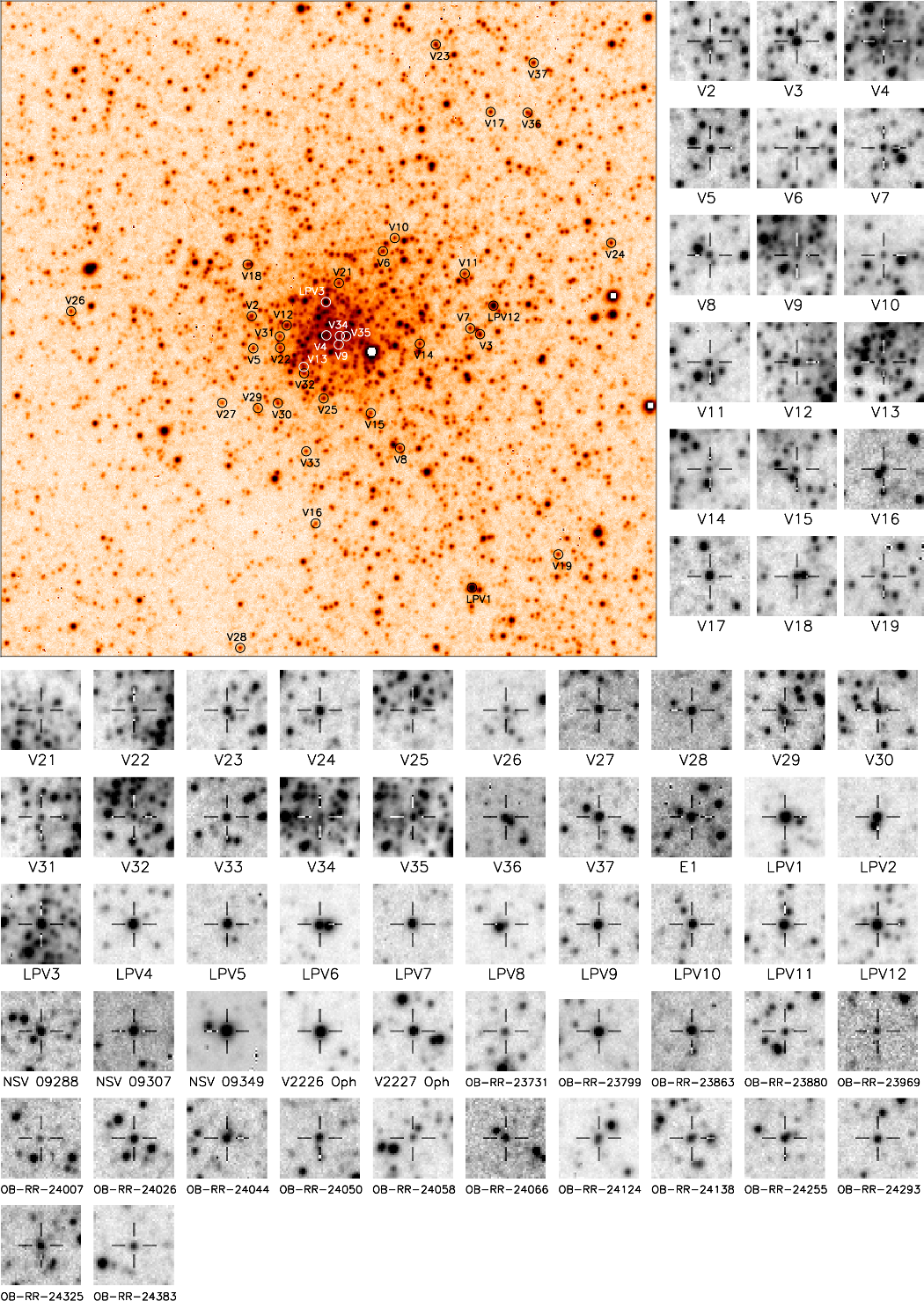}
 \caption{Finding chart constructed from our SDSS-$g\arcmin$ band reference image covering the cluster region. North is up and East is to the right. The cluster image is 4.8$\times$4.8 arcmin$^{2}$. All of the confirmed variables in this sub-field are identified. The image stamps are of size 19.3$\times$19.3 arcsec$^{2}$. Each confirmed variable lies at the centre of its corresponding image stamp and is marked by a cross-hair. Image stamps are presented for all confirmed variables in our full FoV. Note that the image stamp for OB-RR-24066 is taken from our SDSS-$i\arcmin$ band reference image instead, since it is not detected in the SDSS-$g\arcmin$ band reference image. OGLE variable names have been abbreviated for clarity.}
 \label{fchart_centre}
\end{figure*}

\subsection{Detection of new variable stars}
Our pipeline generated 26727 and 31762 light curves for the stars measured in the SDSS-$g\arcmin$ and SDSS-$i\arcmin$ reference images respectively. 
We proceeded by compiling a list of known variables in a $16\arcmin \times 16\arcmin$ area centred on NGC~6401 using the equatorial coordinates and finding charts available in the OGLE-IV database\footnote{ftp://ftp.astrouw.edu.pl/ogle/ogle4/OCVS/blg/rrlyr/} and the coordinates provided by \citet{Samus09}. We also considered long period variables in the General Catalog of Variable Stars (GCVS) \footnote{http://www.sai.msu.su/gcvs/gcvs/index.htm} \citep{Samus09_2} and the International Variable Star Index (VSX) of the American Association of Variable Star Observers (AAVSO) \footnote{https://www.aavso.org/vsx/}.

In order to identify variable sources in our data, we employed three different methods. Using all SDSS-$g\arcmin$ band light curves, we performed a Lomb-Scargle (LS) periodogram search for periods between 0.1 to 1.3 days (method 1), retaining the best-fit periods. A subsequent second run with no restriction on the period was performed to identify variables that vary on longer timescales. For the same set of light curves, we also evaluated the ${\cal S_{B}}$ variability statistic (method 2), as defined in equation 3 from \citet{Figuera13}. Data points deviating more than three times the RMS error from the mean magnitude were clipped during this process. This removed between 0-4 data points from the light curves but mostly left them unaffected.
Lastly, for each filter we constructed a `residual superimage' by taking the absolute value of the residual ADU counts at each pixel ($i,j$) for every difference image and summing over the images (method 3). For a detailed description of the method, see \citet{Bramich11}. Variable sources were easier to spot with this method since their PSF-like peaks on the superimage were strongly enhanced.

To select candidate variable stars, we required ${\cal S_{B}}>$1.0 and best-period LS Power score$>$0.4. These selection criteria were decided empirically by looking at the respective distributions of these parameters, and were intentionally set low enough so as to ensure no candidate variables were missed.

Upon visual inspection of the candidates, after removing duplicates and bad quality light curves, 62 were retained as showing genuine variability. Almost all previously known variables were identified by this procedure. However, we did not find any evidence of variability for any of the sources identified in the immediate vicinity of the coordinates of previously reported variables V1, V20, which were also not found in the OGLE variable database, and OGLE-BLG-RRLYR-23724. We note that the OGLE database also does not contain any variables in the reported coordinates of V3, V8, V10 and V14, all of which were identified by our analysis.

From the residuals superimages in each filter, we selected candidate variables by visually identifying all PSF-like objects. We then examined the light curves corresponding to the coordinates of these objects and identified 8 additional variables, bringing the total to 70. Our results are summarized in tables~\ref{tab:var} and \ref{tab:var_out}, which list all of the confirmed variable stars in our field of view.
\begin{figure*}
  \begin{tabular}{@{}cccc@{}}
    \includegraphics[width=.23\textwidth]{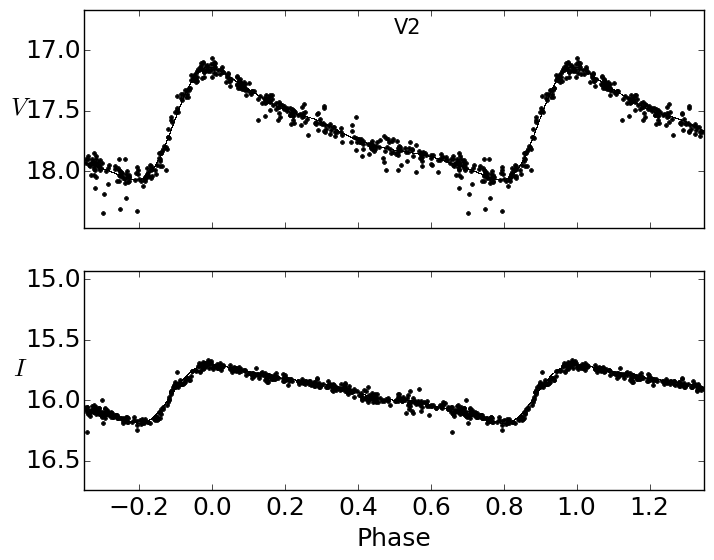} & 
    \includegraphics[width=.235\textwidth]{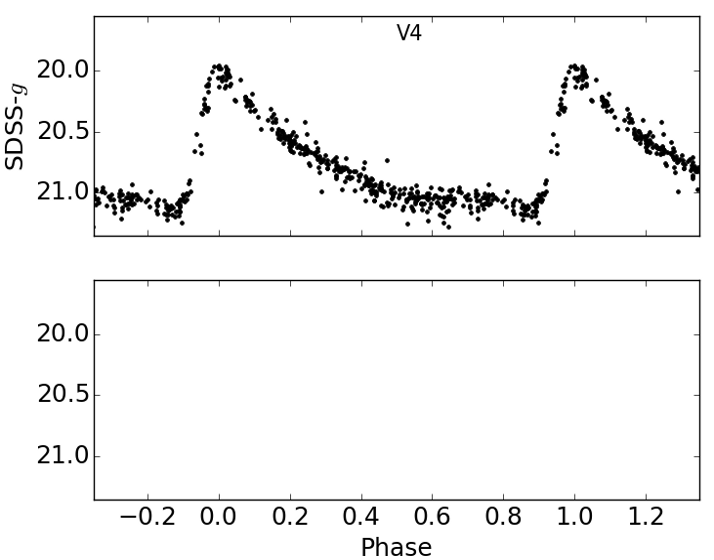} &
    \includegraphics[width=.23\textwidth]{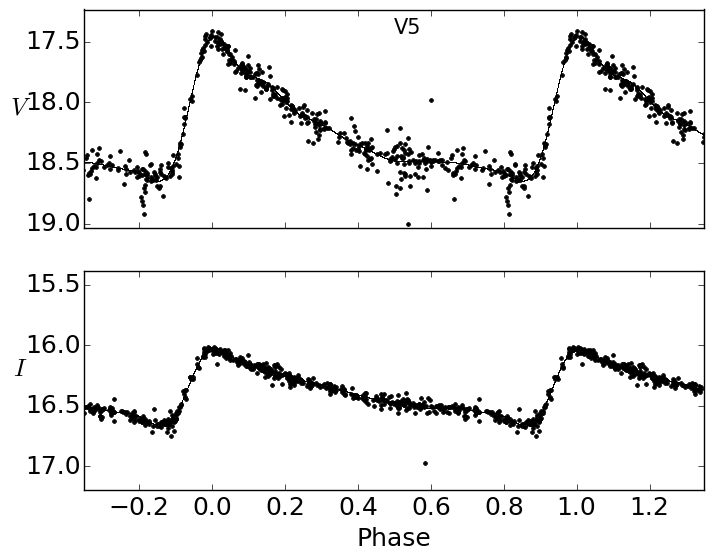} &
    \includegraphics[width=.23\textwidth]{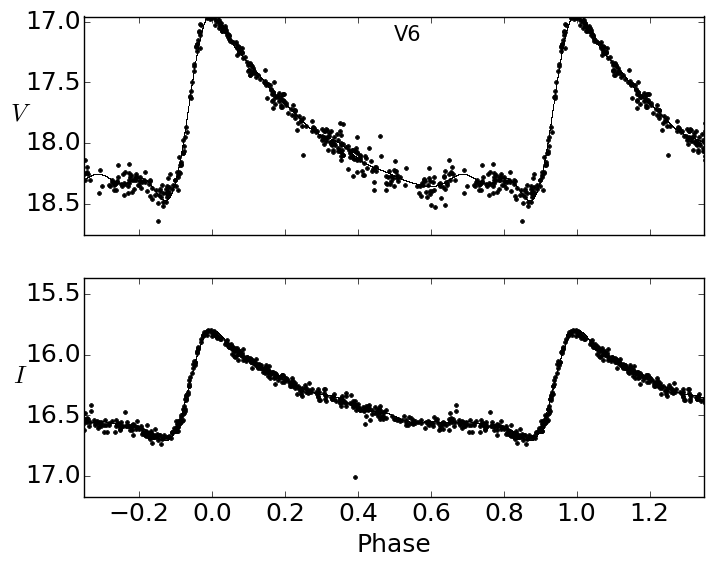} \\
    \includegraphics[width=.23\textwidth]{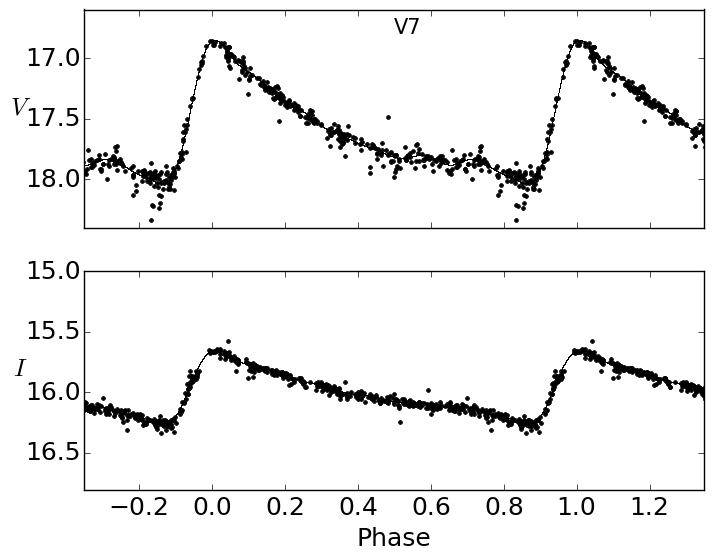} & 
    \includegraphics[width=.23\textwidth]{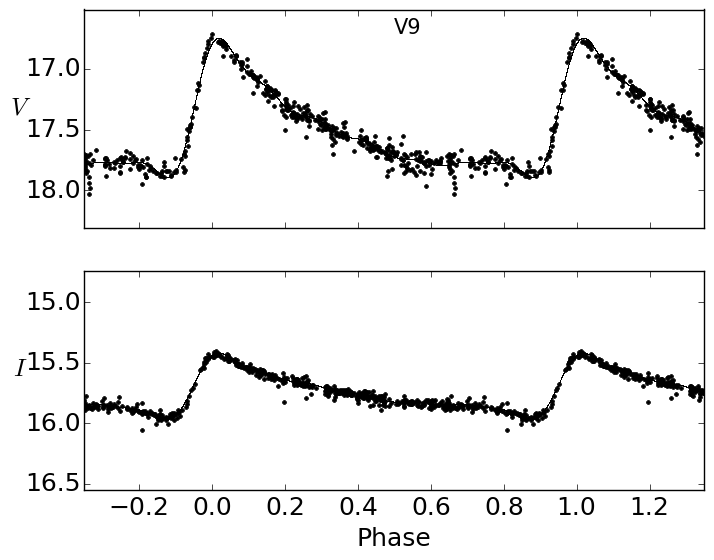} &
    \includegraphics[width=.23\textwidth]{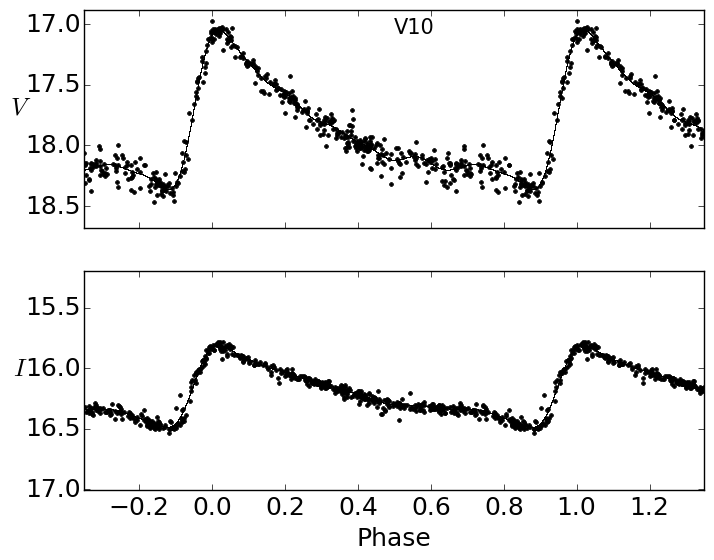} &
    \includegraphics[width=.23\textwidth]{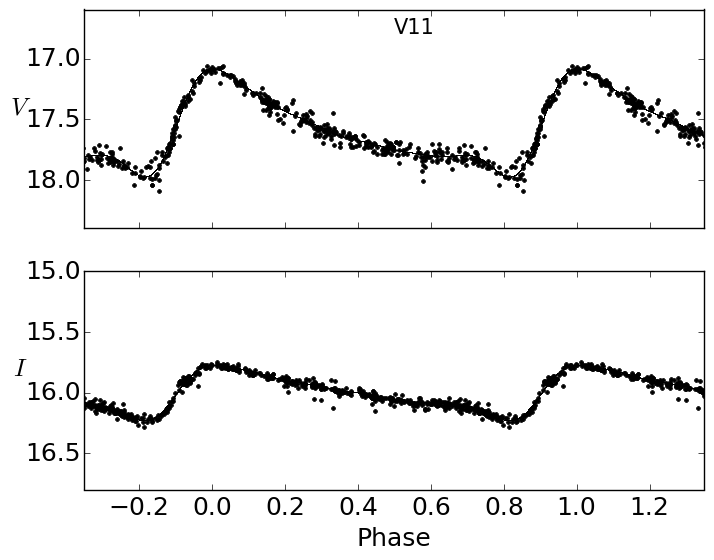} \\
    \includegraphics[width=.23\textwidth]{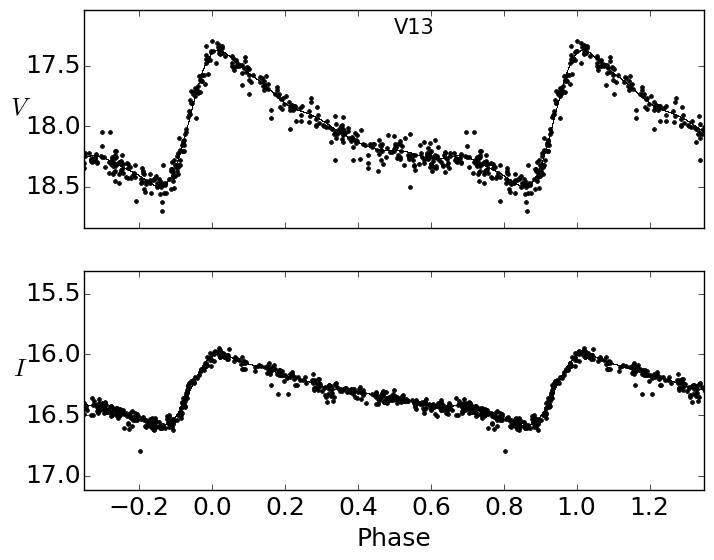} & 
    \includegraphics[width=.23\textwidth]{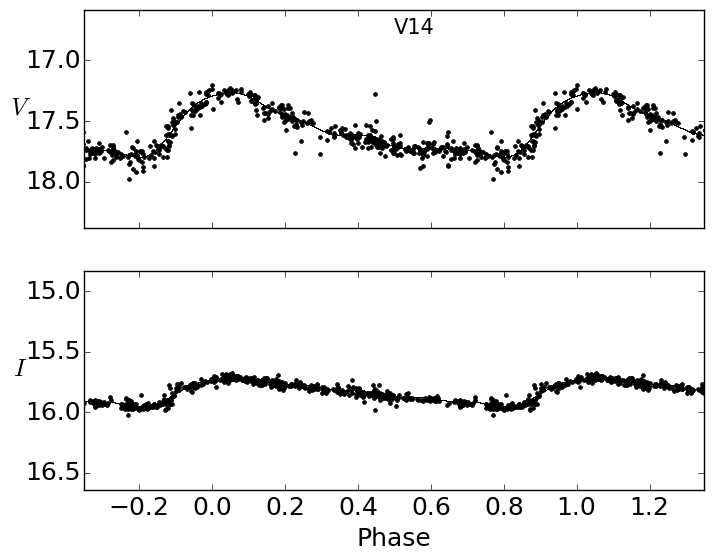} &
    \includegraphics[width=.23\textwidth]{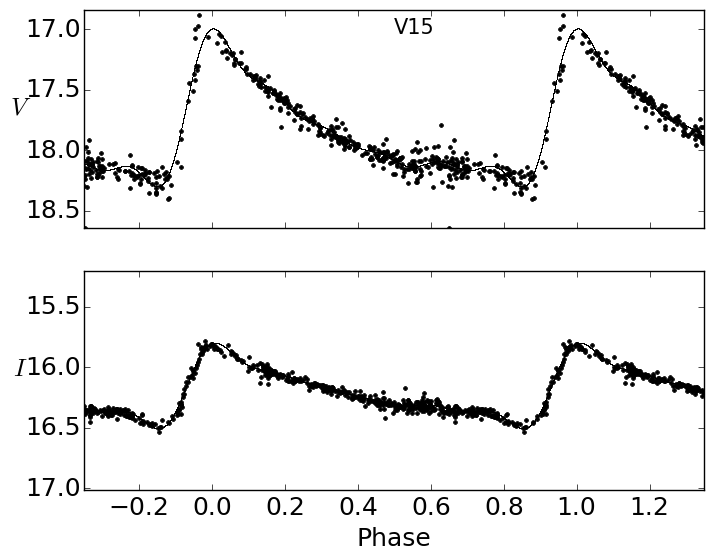} &
    \includegraphics[width=.23\textwidth]{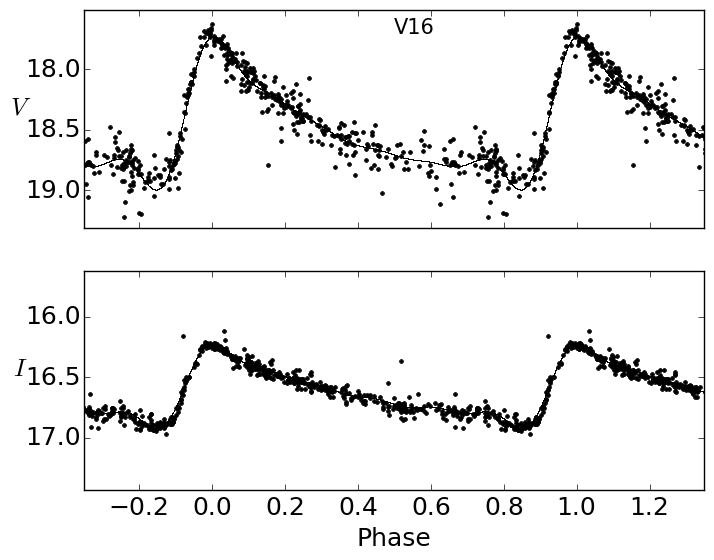} \\
    \includegraphics[width=.23\textwidth]{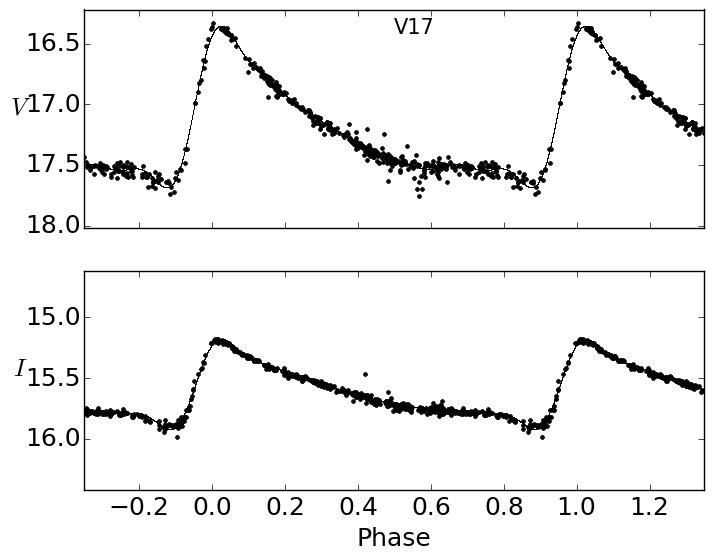} & 
    \includegraphics[width=.23\textwidth]{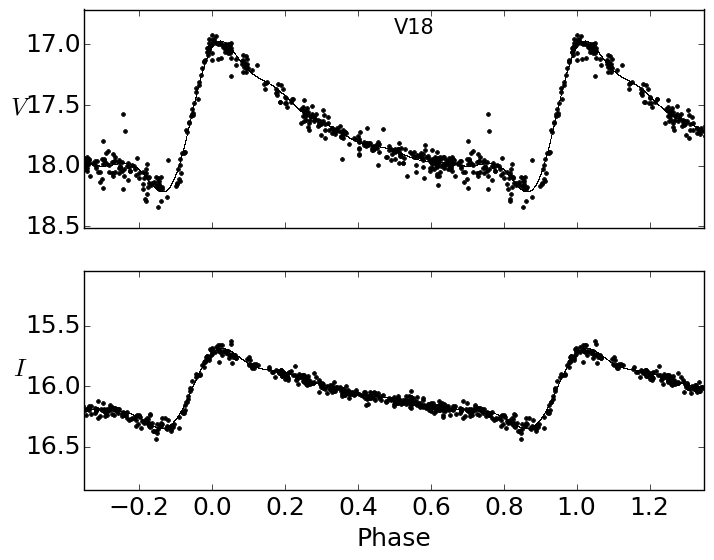} &
    \includegraphics[width=.23\textwidth]{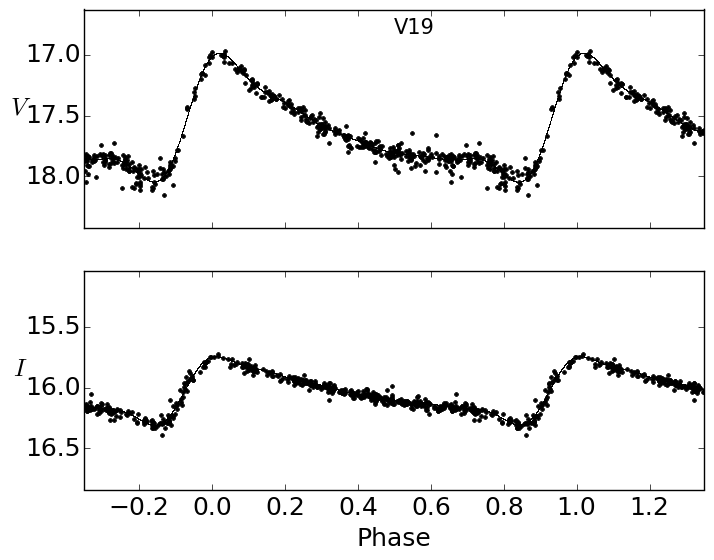} &
    \includegraphics[width=.23\textwidth]{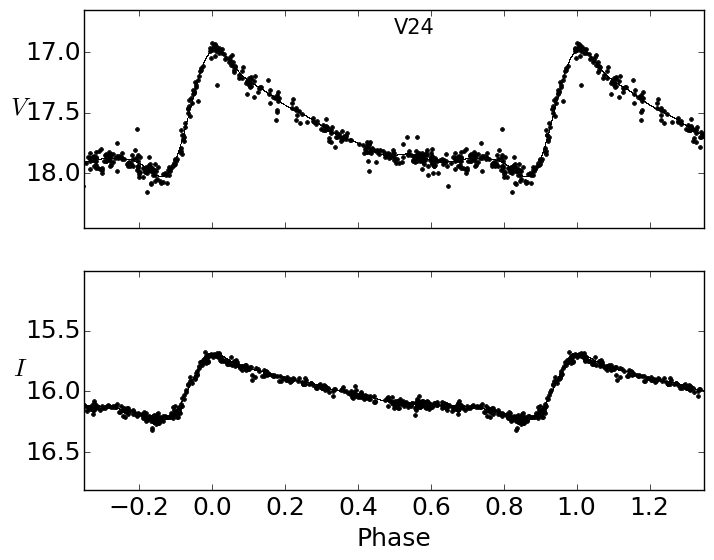} \\
    \includegraphics[width=.23\textwidth]{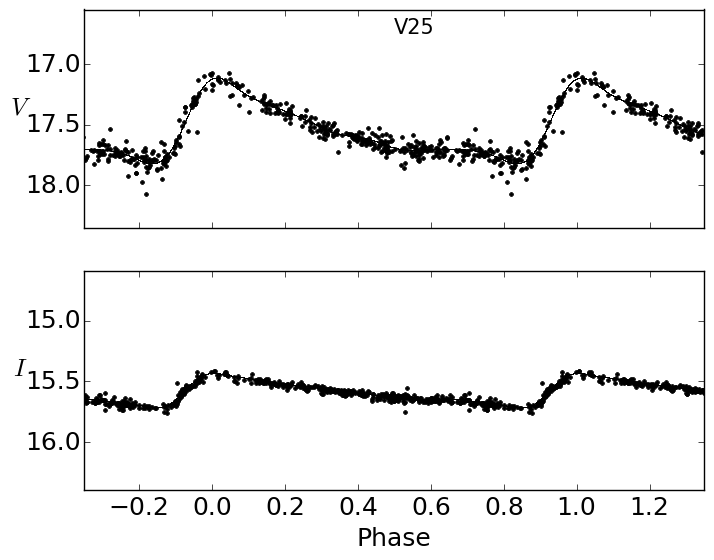} & 
    \includegraphics[width=.23\textwidth]{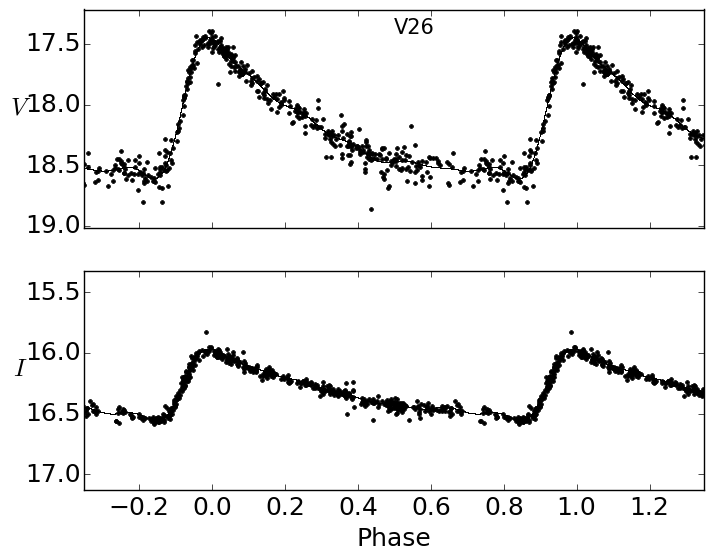} &
    \includegraphics[width=.23\textwidth]{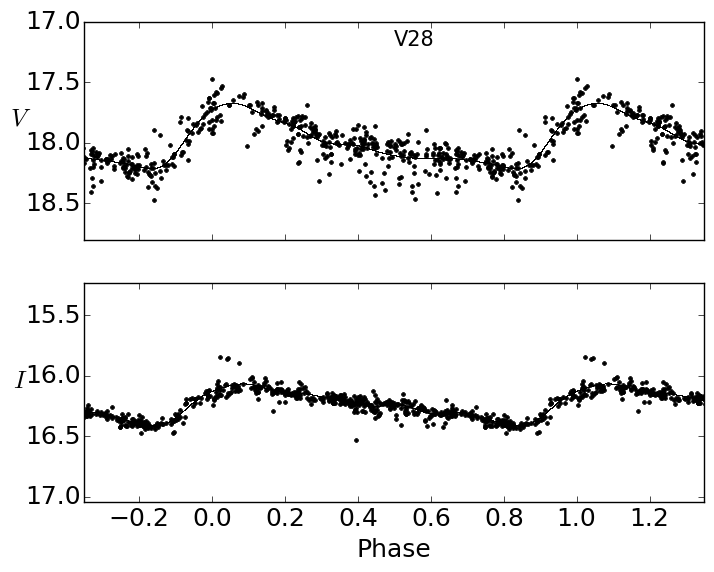} &
    \includegraphics[width=.23\textwidth]{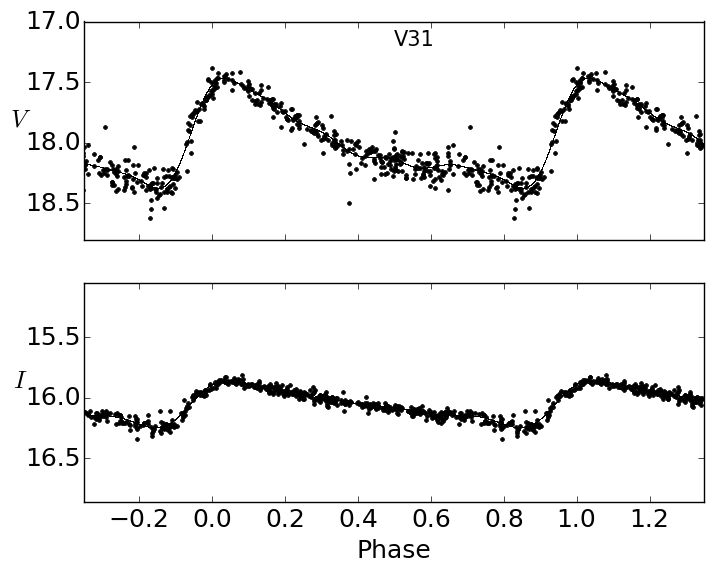} \\
    \includegraphics[width=.23\textwidth]{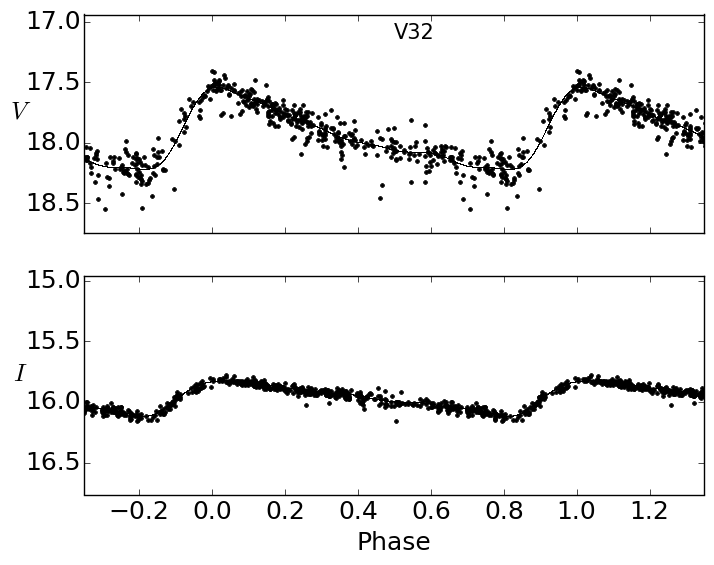} & 
    \includegraphics[width=.23\textwidth]{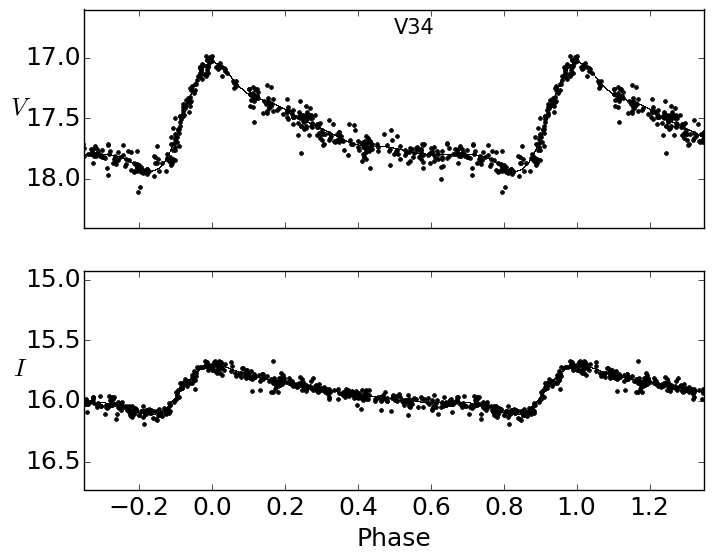} &
    \includegraphics[width=.23\textwidth]{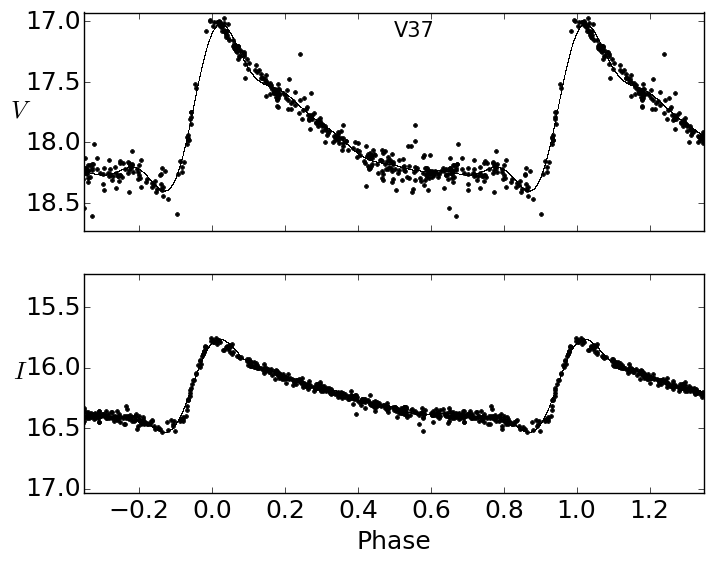} &
  \end{tabular}
  \caption{Standard SDSS-$g\arcmin$ and SDSS-$i\arcmin$ band light curves, linearly transformed to the $V$ and $I$ bands respectively, for the RRab stars in NGC 6401 phased with the periods listed in column 5 of Table~\ref{tab:var} (or column 4 if no OGLE period was available). Note that no SDSS-$i\arcmin$ band data were available for V4 due to its proximity to a bad pixel region on the CCD so only SDSS-$g\arcmin$ band instrumental magnitudes are displayed.}
  \label{fig:rrab}
\end{figure*}

\begin{figure*}
  \begin{tabular}{@{}cccc@{}}
    \includegraphics[width=.23\textwidth]{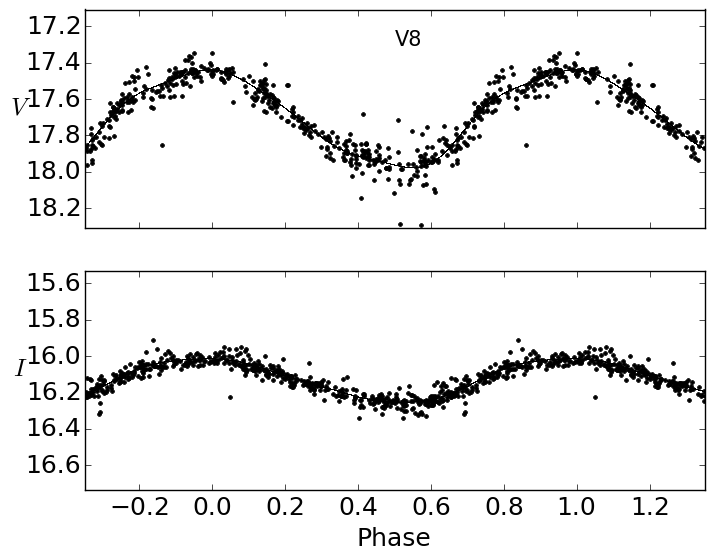} &
    \includegraphics[width=.23\textwidth]{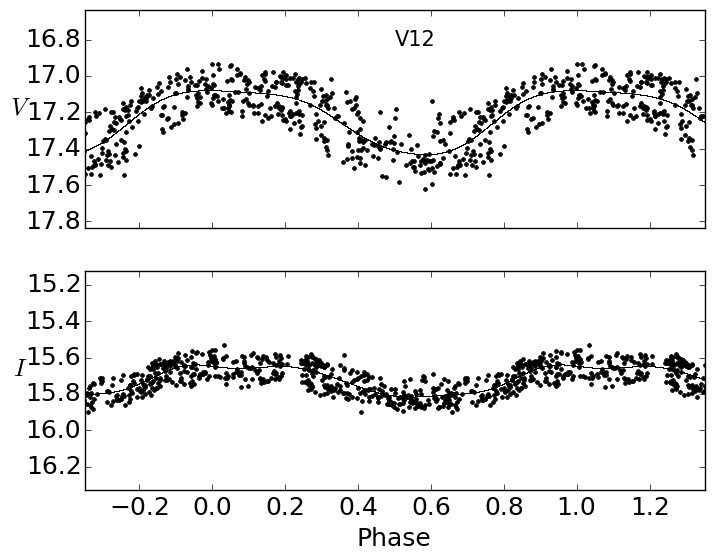} &
    \includegraphics[width=.23\textwidth]{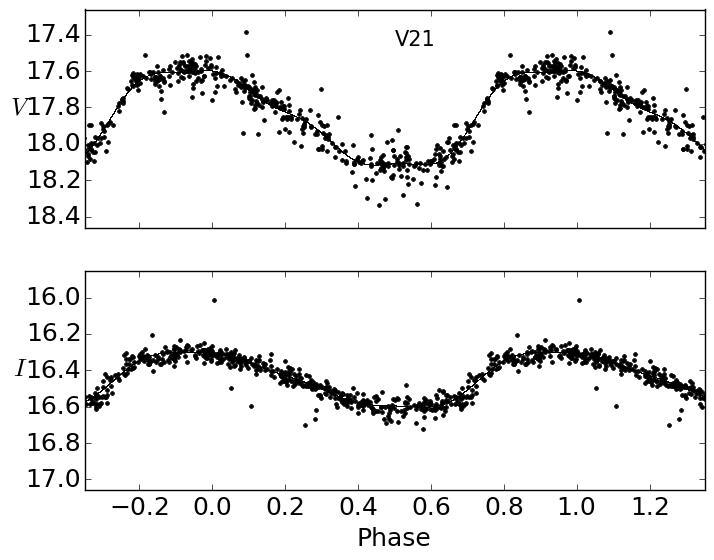} &
    \includegraphics[width=.23\textwidth]{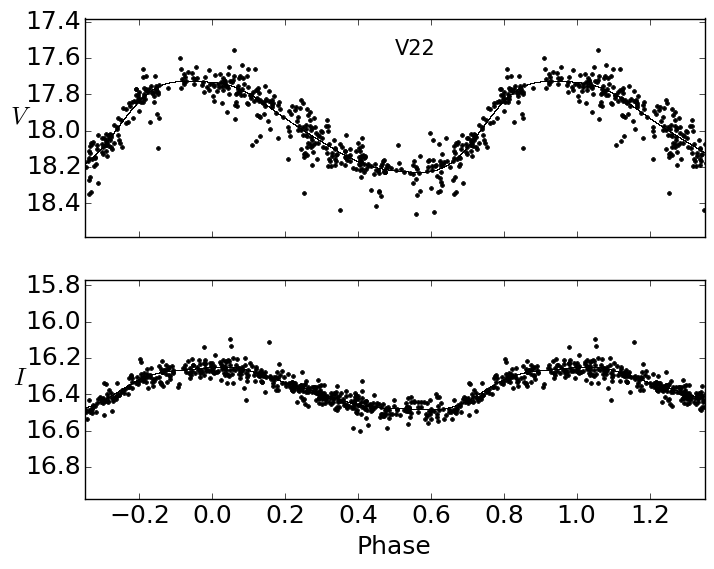} \\
    \includegraphics[width=.23\textwidth]{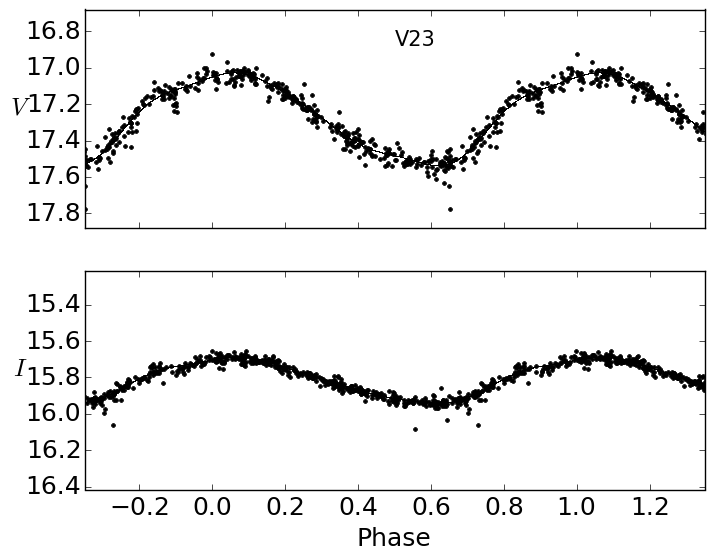} & 
    \includegraphics[width=.23\textwidth]{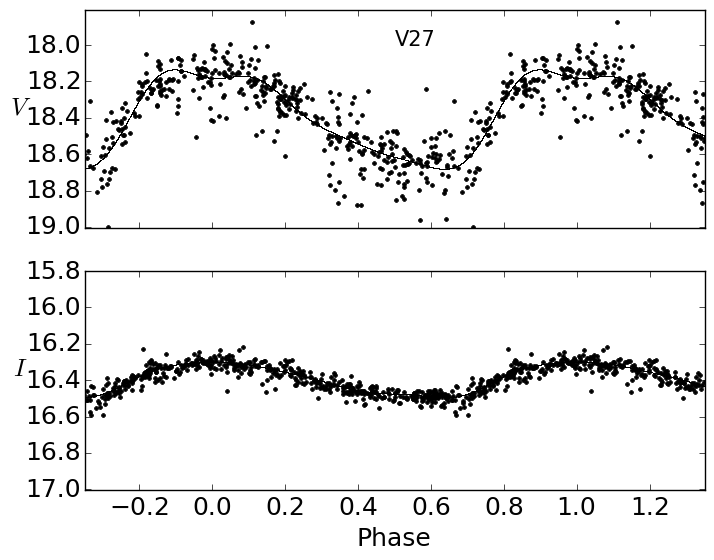} &
    \includegraphics[width=.23\textwidth]{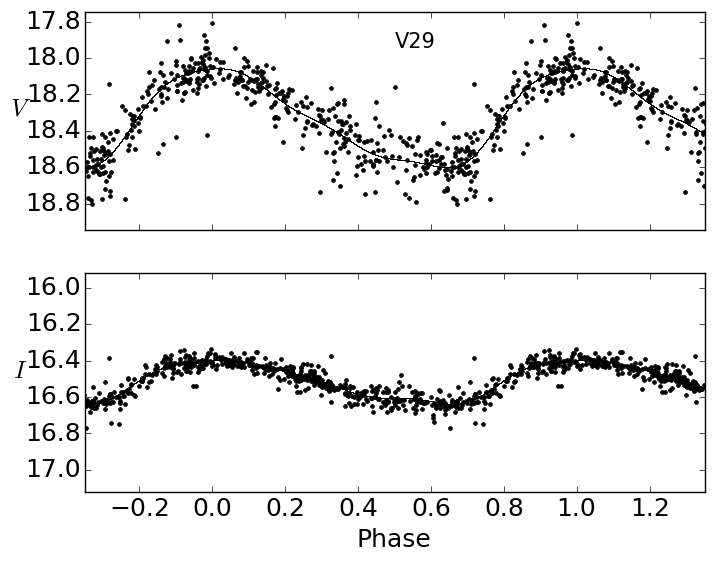} &
    \includegraphics[width=.23\textwidth]{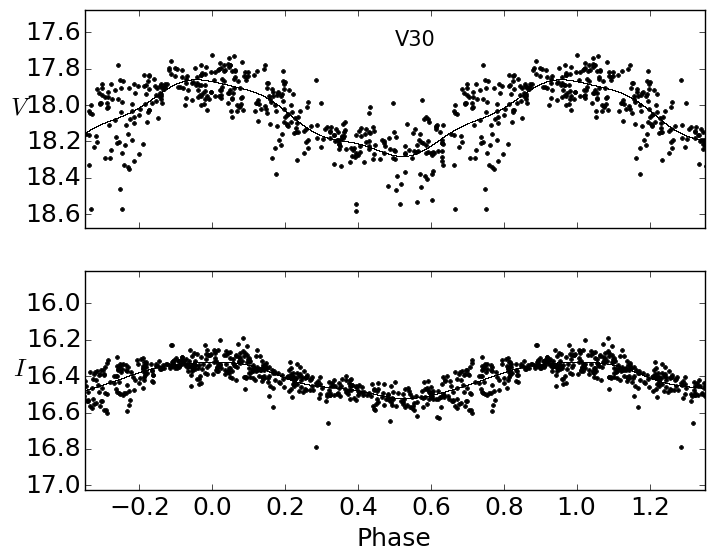} \\
    \includegraphics[width=.23\textwidth]{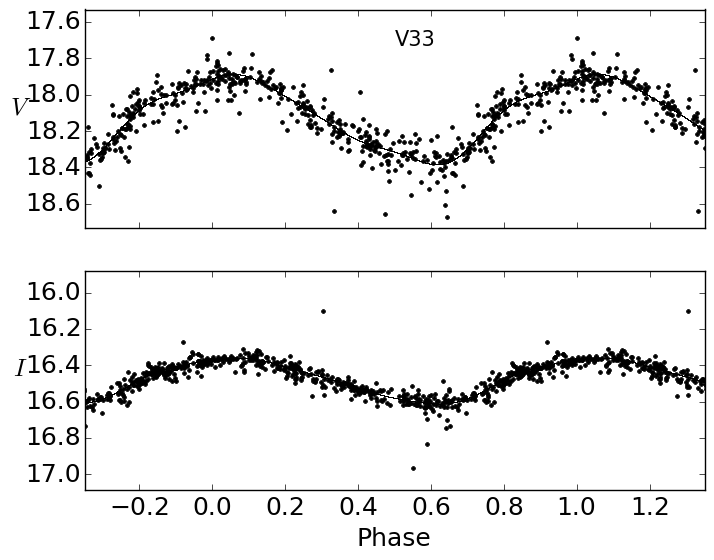} &
    \includegraphics[width=.23\textwidth]{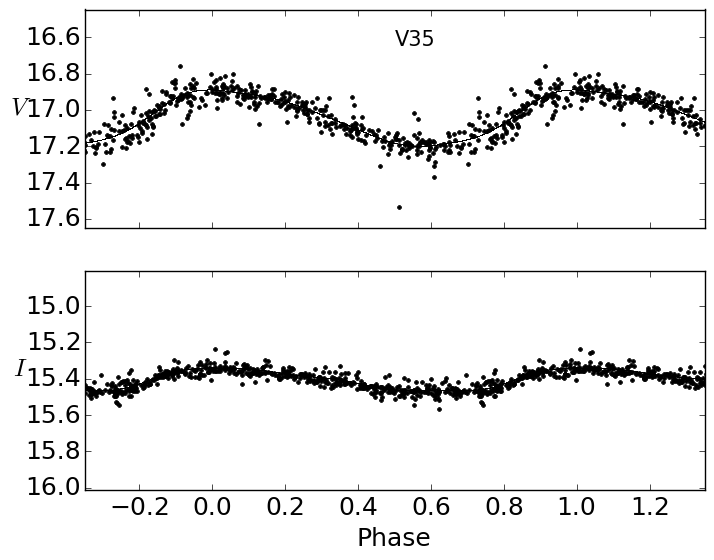} & 
    \includegraphics[width=.23\textwidth]{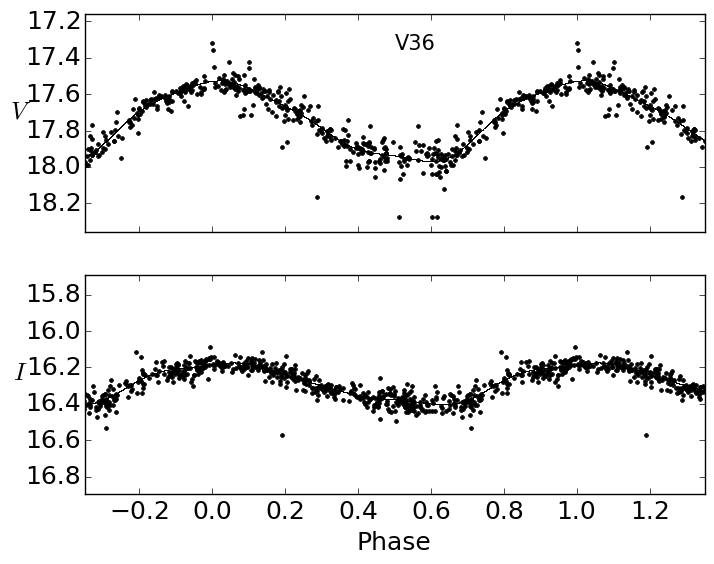} &                                                                                                    
  \end{tabular}
  \caption{Standard SDSS-$g\arcmin$ and SDSS-$i\arcmin$ band light curves, linearly transformed to the $V$ and $I$ bands respectively, for the RRc stars in NGC 6401 phased with the periods listed in column 5 of Table~\ref{tab:var} (or column 4 if no OGLE period was available).}
  \label{fig:rrc}
\end{figure*}

We found 37 RRab and 14 RRc variables within a 15.8 $\times$ 15.8 arcmin field centred at the coordinates of NGC~6401. Of these, 23 RRab and 11 RRc are within one cluster radius\footnote{The list of all 158 globular clusters in the Milky Way is available at http://messier.obspm.fr/xtra/supp/mw\_gc.html} (2.4\arcmin) from the centre of NGC~6401 and most likely belong to the cluster. Finding charts are provided in Figures~\ref{fchart_full} and \ref{fchart_centre}. The variable light curves for the likely cluster members are displayed in Figures~\ref{fig:rrab} and \ref{fig:rrc}, and their locations on the CMD are shown in Figure~\ref{cmd}. The light curves of other variables are plotted in Figures~\ref{fig:varout1} and \ref{fig:varout2}.

Cluster variables listed in table~\ref{tab:var} are denoted by the letter V, with all numbers up to 25 belonging to the Terzan \& Rutily lists. OGLE identifiers, where available, are also provided in the second column. Our best fit periods and the OGLE periods are given in columns four and five respectively. Non-member variables are listed in Table~\ref{tab:var_out}, which includes 17 long period variables (LPV) with periods $>27$ days. Their light curves are displayed in the Appendix. Long period variables (LPV) 1 to 12 are new discoveries. We note that the shape of the light curve of variable V3 suggests that it is most likely a W Virginis (CWB) star with a period of $\sim$1.75 days. The eclisping binary E1 is also a new discovery. All light curves are available for download in the electronic version of this article. An excerpt from the full table of electronic data is provided in Table~\ref{tab:gi_phot}.

\begin{table*}
\scriptsize
\begin{center}
\caption[] {\small Fourier decomposition parameters ($V$-band).}
\hspace{0.01cm}
 \begin{tabular}{lccccccccccc}
\hline 
Star & $A_0$ & $A_1$ & $A_2$ & $A_3$ & $A_4$ & $\phi_{21}$ &  $\phi_{31}$ &  $\phi_{41}$ & N & $D_m$  \\
\hline
 RRab stars & & & & & & & & & &\\
\hline
 V2  & 17.664 & 0.271 & 0.128 & 0.080 & 0.032 & 4.440 & 9.069 & 7.248 & 6 & 3.14 \\ 
 V5  & 18.234 & 0.323 & 0.160 & 0.111 & 0.065 & 3.953 & 8.313 & 6.350 & 7 & 1.52 \\ 
 V6  & 17.956 & 0.424 & 0.201 & 0.156 & 0.099 & 3.821 & 7.935 & 5.931 & 8 & 3.08 \\ 
 V7  & 17.598 & 0.307 & 0.165 & 0.110 & 0.072 & 3.984 & 8.391 & 6.479 & 8 & 2.39 \\ 
 V9  & 17.510 & 0.289 & 0.142 & 0.109 & 0.070 & 3.950 & 8.089 & 6.053 & 7 & 1.70 \\ 
 V10 & 17.878 & 0.344 & 0.176 & 0.124 & 0.083 & 3.972 & 8.323 & 6.321 & 9 & 2.86 \\ 
 V11 & 17.593 & 0.243 & 0.124 & 0.079 & 0.045 & 4.153 & 8.603 & 6.964 & 6 & 1.13 \\ 
 V13 & 18.037 & 0.292 & 0.163 & 0.105 & 0.064 & 4.102 & 8.596 & 6.779 & 7 & 3.18 \\ 
 V14 & 17.579 & 0.164 & 0.070 & 0.033 & 0.015 & 4.135 & 8.446 & 7.417 & 6 & 7.88 \\ 
 V15 & 17.839 & 0.336 & 0.172 & 0.123 & 0.078 & 3.948 & 8.221 & 6.197 & 6 & 1.81 \\ 
 V16 & 18.508 & 0.320 & 0.166 & 0.117 & 0.088 & 4.006 & 8.282 & 6.327 & 6 & 2.30 \\ 
 V17 & 17.217 & 0.358 & 0.174 & 0.130 & 0.081 & 3.901 & 8.158 & 6.143 & 7 & 1.01 \\ 
 V18 & 17.712 & 0.316 & 0.170 & 0.110 & 0.080 & 4.081 & 8.429 & 6.636 & 6 & 1.94 \\ 
 V19 & 17.625 & 0.282 & 0.147 & 0.097 & 0.060 & 4.065 & 8.495 & 6.692 & 6 & 1.15 \\ 
 V24 & 17.647 & 0.277 & 0.136 & 0.096 & 0.067 & 3.908 & 8.241 & 6.213 & 7 & 2.21 \\ 
 V25 & 17.549 & 0.184 & 0.090 & 0.061 & 0.035 & 4.009 & 8.423 & 6.678 & 7 & 3.41 \\ 
 V26 & 18.216 & 0.335 & 0.150 & 0.103 & 0.065 & 3.936 & 8.229 & 6.051 & 7 & 2.63 \\ 
 V28 & 17.999 & 0.170 & 0.082 & 0.041 & 0.018 & 4.144 & 8.607 & 7.232 & 7 & 5.37 \\ 
 V31 & 18.000 & 0.246 & 0.126 & 0.072 & 0.043 & 4.033 & 8.574 & 6.856 & 6 & 4.73 \\ 
 V32 & 17.941 & 0.201 & 0.094 & 0.047 & 0.020 & 4.328 & 8.983 & 7.103 & 6 & 2.72 \\ 
 V34 & 17.603 & 0.227 & 0.117 & 0.077 & 0.056 & 3.964 & 8.370 & 6.413 & 6 & 2.62 \\ 
 V37 & 17.928 & 0.370 & 0.173 & 0.126 & 0.096 & 3.844 & 8.138 & 6.007 & 6 & 1.89 \\ 
\hline
 RRc stars & & & & & & & & & & \\
\hline
 V8  & 17.707 & 0.205 & 0.015 & 0.013 & 0.007 & 5.002 & 4.222 & 2.888 & 4 & --- \\ 
 V12 & 17.234 & 0.144 & 0.024 & 0.012 & 0.001 & 5.968 & 3.033 & 1.745 & 4 & --- \\ 
 V21 & 17.861 & 0.226 & 0.035 & 0.015 & 0.020 & 4.758 & 2.985 & 1.830 & 4 & --- \\ 
 V22 & 17.983 & 0.202 & 0.026 & 0.015 & 0.009 & 4.586 & 3.359 & 1.670 & 4 & --- \\ 
 V23 & 17.279 & 0.196 & 0.015 & 0.016 & 0.010 & 5.029 & 4.357 & 3.074 & 4 & --- \\ 
 V27 & 18.404 & 0.195 & 0.037 & 0.030 & 0.013 & 4.689 & 3.543 & 1.477 & 4 & --- \\ 
 V29 & 18.344 & 0.205 & 0.039 & 0.015 & 0.014 & 4.408 & 3.193 & 2.263 & 4 & --- \\ 
 V30 & 18.074 & 0.156 & 0.007 & 0.008 & 0.011 & 3.731 & 4.977 & 5.157 & 4 & --- \\ 
 V33 & 18.131 & 0.185 & 0.018 & 0.016 & 0.010 & 5.075 & 4.747 & 3.342 & 4 & --- \\ 
 V35 & 17.048 & 0.117 & 0.015 & 0.008 & 0.005 & 4.808 & 2.378 & 1.185 & 4 & --- \\ 
 V36 & 17.756 & 0.174 & 0.016 & 0.012 & 0.009 & 4.477 & 3.908 & 3.005 & 4 & --- \\ 
\hline
\label{tab:fourier_coeffs}
\end{tabular}
\end{center}
\end{table*}

\subsection{The case of variable V12}
\label{sec:V12}
\begin{figure*}
  \begin{tabular}{@{}cc@{}}
    \includegraphics[width=.46\textwidth]{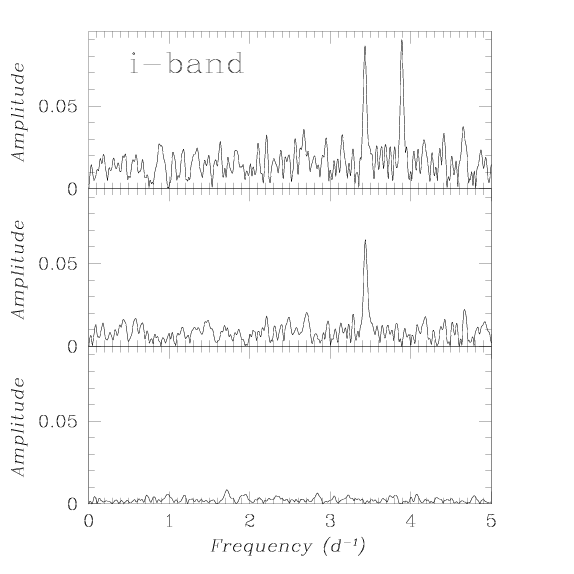} & 
    \includegraphics[width=.46\textwidth]{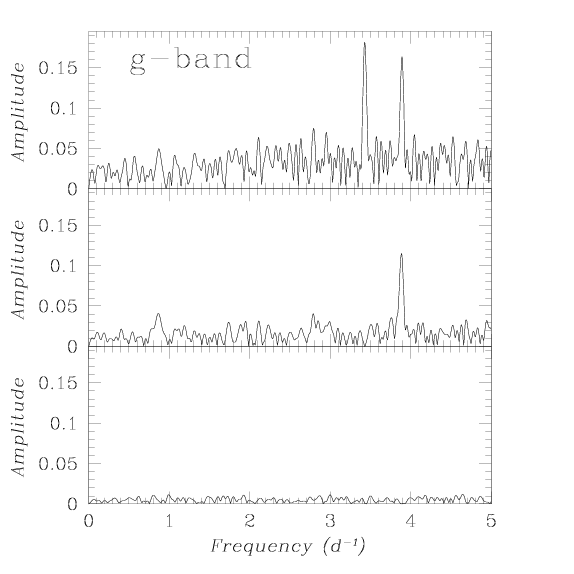} \\                                                                                             
  \end{tabular}
  \caption{Frequency spectra for V12 in the SDSS-$i\arcmin$ (left) and SDSS-$g\arcmin$ (right) bands.}
  \label{fig:V12}
\end{figure*}

We detected two independent frequencies in the  SDSS-$g\arcmin$ and SDSS-$i\arcmin$ light curves of the
RR Lyrae. The frequency spectra are shown in Figure~\ref{fig:V12}.
The top panels show the prominent peaks, which alternate in intensity between the g-band and
the i-band case. Pre-whitening the strongest peak leads to the spectra shown in the middle
panels. In both cases the secondary peak persists, which demonstrates that these signals
are not aliases of each other. Subsequent pre-whitening of the secondary peak removes
all signals as seen in the bottom panels.

The frequency ratio of these two signals is 0.884, i.e., far from the canonical 0.745
expected if this star was a radial double mode pulsator RR01 (RRd) (fundamental-first
overtone) or 0.80 for RR12 stars (first and second overtone modes). No RR12
have yet been identified in globular clusters but three were detected by \citet{Alcock00} in 
the Large Magellanic Cloud (LMC). From this we may conclude that in the case of V12 either one or both
pulsation modes are non-radial. 

Other non-radial pulsators among first overtone or RR1 RR Lyraes have been identified in the globular cluster M55 \citep{Olech99} and by \citep{Alcock00} in the LMC.
The fundamental difference of V12 with all these non-radial pulsator RR Lyrae is that
in all of them the frequency ratio is always larger than 0.95, whereas it is only 0.884 in V12. 
Thus V12 adds to the list of non-radial pulsator RR Lyrae stars in globular clusters. Precise determination of the excited non-radial mode would require abundant and accurate data.

\section{Determination of the physical parameters from the RR Lyr variables}
\label{sec:Physical}
We used Fourier decomposition of the calibrated SDSS-$g\arcmin$ and SDSS-$i\arcmin$ band RR Lyr light curves into their harmonics in order to estimate the metallicity, luminosity and effective temperature of the stars. Fourier decomposition involves modeling the light curves as a Fourier series of the form:

\begin{equation}
m(t) = A_0 ~+~ \sum_{k=1}^{N}{A_k ~\cos~\left[ {2\pi \over P}~k~(t-E) ~+~ \phi_k \right] },
\label{eq_foufit}
\end{equation}

\noindent
where $m(t)$ is the magnitude at time $t$, $N$ the number of harmonics used in the fit, $P$ the period of variability, and $E$ the reference epoch. $A_k$ and $\phi_k$ are the amplitudes and phases of the sinusoidal components respectively. The fit to the data was performed using linear least squares, and the best fit values of the amplitudes and phases of the harmonics were evaluated. From these, the Fourier parameters $\phi_{ij} = j\phi_{i} - i\phi_{j}$ and $R_{ij} = A_{i}/A_{j}$ were determined. In order to avoid fitting noise in the data, each variable was modeled with the lowest possible number of harmonics. We note that, where available, we used the OGLE data to refine the estimates of the periods during the fitting process. 

Since our observations were originally in the SDSS-$g\arcmin$ and SDSS-$i\arcmin$ bands, we performed our fits in each band using the alternative Fourier form: 
\begin{equation}
m(t) = A_0 + \sum_{k=1}^{N}{B_k \cos\left[ {2\pi \over P}k(t-E)\right]\\
           + C_k\sin\left[ {2\pi \over P}k(t-E)\right] },
\label{eq_foufit}
\end{equation}
from which we derived the coefficients $B_k$ and $C_k$. These Fourier coefficients were then converted to their equivalents in the $V$-band using 
the \citet{Lupton05} transformations between SDSS magnitudes and UBVRcIc combined with our equation~\ref{eqn:istd}, from which we derived $B_k(V) =  0.5778 B_k(g) + 0.4222 B_k(i)$ and $C_k(V) = 0.5778 C_k(g) + 0.4222 C_k(i)$. We then estimated the coefficients $A_k(V)=\sqrt{B_k(V)^2 + C_k(V)^2}$ and $\phi_k(V)=\cos^{-1}(B_k(V)/A_k(V))$.

The best fit parameters and number of harmonics used for individual RRab and RRc variable light curves are listed in Table~\ref{tab:fourier_coeffs}. Some variables did not provide useful constraints and were excluded from further analysis. Specifically, when determining the physical parameters, we did not use the following variables: V12  exhibits non-radial pulsation (see section~\ref{sec:V12}); V30 shows too much scatter and its Fourier parameters are unreliable; V14 and V28 have a {\it deviation parameter}, defined by \citet{Jurcsik96}, $D_{\rm m} > 5.0$ and are therefore not suitable to use for calibration purposes. V4 was also not used because it has no SDSS-$i\arcmin$ data.

\subsection{Metallicity}
The Fourier decomposition parameters of RRab stars can be used to calculate the metallicity from the semi-empirical relationships of Jurcsik \& Kov\'acs (1996), which express [Fe/H] as a function of the period and Fourier parameter $\phi^{(s)}_{31}$, where the superscript $s$ denotes a Fourier {\it sine} series. Since we used a {\it cosine} series for our fits (\ref{eq_foufit}), we converted to {\it sine} using $\phi^{(s)}_{ij} = \phi_{ij} - (i-j)\frac{\pi}{2}$.

The metallicity is then given by:
\begin{equation}
{\rm [Fe/H]}_{\mathrm{J}} = -5.038 ~-~ 5.394~P ~+~ 1.345~\phi^{(s)}_{31},
\label{eq:JK96}
\end{equation}
\noindent where the subscript $J$ denotes a non-calibrated metallicity. This can be converted to the metallicity scale of \citet{Zinn84} (hereafter ZW) using the following relationship from \citet{Jurcsik95}:
\begin{equation}
{\rm [Fe/H]}_{\mathrm{ZW}} = \frac{{\rm [Fe/H]}_J - 0.88}{1.431}.
\end{equation}

Equation~\ref{eq:JK96} is only applicable to RRab stars with a {\it deviation parameter} $D_m$ below a given limit. Although \citet{Jurcsik96} originally used $D_{\rm m} < 3.0$, some authors have relaxed that condition to $D_{\rm m} < 5.0$ with the aim of improving the statistics of the mean physical parameters. We therefore adopted $D_{\rm m} < 5.0$ as a selection criterion to estimate stellar properties for our RRab stars and have rejected stars V14 and V28 with  $D_{\rm m} > 5.0$ when estimating the physical parameters.

The metallicity can also be obtained for the RRc stars using 
${\rm [Fe/H]}_{\mathrm{ZW}} = 52.466~P^2 ~-~ 30.075~P ~+~ 0.131~\phi_{31}^{2}$
\begin{equation}
~~~~~~~	~+~ 0.982 ~ \phi_{31} ~-~ 4.198~\phi_{31}~P ~+~ 2.424,
\label{eq:Morgan07}
\end{equation}
\noindent from \citet{Morgan07} (her equation 3).

Metallicity values calculated using these relationships are reported in Tables~\ref{tab:phys_rrab} and \ref{tab:phys_rrc}. The transformation to the UVES \citep{Carretta09} scale is given by:
\begin{equation}
{\rm [Fe/H]}_{\mathrm{UVES}} =-0.413 + 0.130 {\rm [Fe/H]}_{\mathrm{ZW}} - 0.356{\rm [Fe/H]}^2_{\mathrm{ZW}}.
\end{equation}

The mean metallicities on the ZW scale for RRab and RRc stars are -1.254 $\pm$ 0.064 and -1.252 $\pm$ 0.210 respectively.

\subsection{Effective Temperature}
The Fourier parameters may also be used to estimate the effective temperature, $T_{\rm eff}$, for RRab and RRc stars.
For the RRab variables, we used the relationships derived by \citet{Jurcsik98}, which link the $(V-K)_0$ colour to the period $P$:
\begin{equation}
\log(T_{\rm eff})= 3.9291 ~-~ 0.1112~(V-K)_0 ~-~ 0.0032~{\rm [Fe/H]},
\end{equation}

\noindent with \\

\noindent
$(V-K)_0 = 1.585 ~+~ 1.257~P ~-~ 0.273~A_1 ~-~ 0.234~\phi^{(s)}_{31}$
\begin{equation}
~+~ 0.062~\phi^{(s)}_{41}.~~~~~~~~~~~~~~~~~~~~~~~~~~~~~~~~~~~~~
\end{equation}

\noindent
For the RRc stars the calibration of \citet{Simon93} can be used:
\begin{equation}
\label{eq:SC93}
\log(T_{\rm eff}) = 3.7746 ~-~ 0.1452~\log(P) ~+~ 0.0056~\phi_{31}.
\end{equation}
Note that the temperatures for RRab and RRc stars calculated using the relationships above are on different absolute scales. For a discussion of the accuracy and caveats of these calibrations we refer the reader to \citet{Cacciari05}, \citet{Ferro08} and \citet{Bramich11}. 

The mean effective temperature for the RRab stars is 6503.87 $\pm$ 78.37 K and for the RRc stars 7397.68 $\pm$ 49.23 K.

\subsection{Absolute Magnitudes}
For the RRab stars, we used \citet{Kovacs01} to derive the absolute magnitude
\begin{equation}
M_{\mathrm{V}} = ~-1.876~\log~P ~-1.158~A_1 ~+0.821~A_3 + K_0,
\label{eq:KW01}
\end{equation}
\noindent where the zero-point of the calibration is $K_0=0.41 \pm 0.02$ mag, as discussed in \citet{Kains15}.

For the RRc stars, equation 10 from \citet{Kovacs98} is used instead, after adopting the zero-point value $K_1=1.061 \pm 0.02$mag used in \citet{Cacciari05}:
\begin{equation}
M_{\mathrm{V}} = K_1 ~-~ 0.961~P ~-~ 0.044~\phi^{(s)}_{21} ~-~ 4.447~A_4.
\label{eq:K98}	
\end{equation}

\noindent The magnitudes were then converted to luminosities using 
\begin{equation}
\log(L/L_{\odot}) = -0.4 (M_{\mathrm{V}} - M_{\rm bol,\odot} + B_C)
\end{equation}
\noindent where $M_{\rm bol,\odot} = 4.75$ is the solar bolometric magnitude and $B_C = 0.06 {\rm [Fe/H]}_{\mathrm{ZW}} + 0.06$ is the bolometric correction term given by \citet{Sandage90}.

The mean absolute magnitudes for the RRab and RRc stars are 0.648 $\pm$ 0.062 and 0.576 $\pm$ 0.032 respectively.

\subsection{Bailey Diagram and Oosterhoff Type}
\label{sec:oost} 
We could not find any assessment of the Oosterhoff type of NGC~6401 in the literature. Using the RR Lyrae variable stars listed in Table~\ref{tab:var}, we estimated mean periods $<P_{\rm RRab}>$=0.55 $\pm$ 0.06 days and $<P_{\rm RRc}>=$ 0.30 $\pm$ 0.03 days. We found that 68\% of the RR Lyrae variables in the cluster are RRab's. These values and the derived metallicity of NGC~6401 indicate that the cluster is Oosterhoff type I (e.g. \citet{Clement01}). 

\begin{figure}
 \centering
 \includegraphics[width=8.0cm]{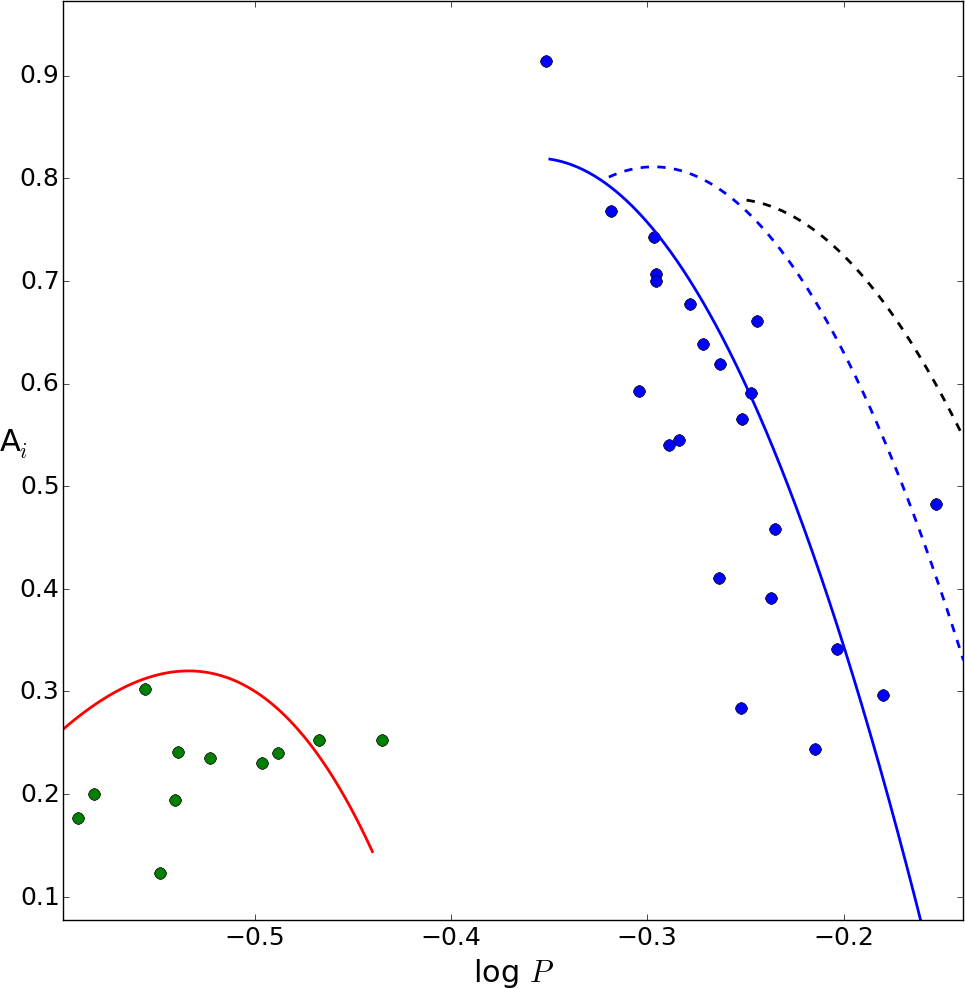}
 \caption{Period-amplitude distributions for the RR Lyrae stars in NGC~6401 in the SDSS-$i\arcmin$ band. RRab and RRc stars are marked with blue and green circles respectively. The dashed black curve was calculated by \citet{Ferro11, Ferro13} for the OoII clusters NGC 5024 and NGC 6333. The solid blue and dashed blue curves are from \citet{Kunder13} for the OoI cluster NGC 2808 and correspond to the mean distributions of regular and evolved stars respectively. The red curve was calculated by \citet{Ferro15} for a sample of RRc stars in five OoI type clusters.}
 \label{fig:bailey}
\end{figure}

The SDSS-$i\arcmin$ band Bailey diagram (log$P$ versus A$_i$) for the cluster RR Lyrae variables is shown in Figure~\ref{fig:bailey}. Since RRab stars have longer periods and greater amplitudes than RRc stars, they occupy distinctly different areas on the figure. Solid and dashed lines mark the mean distributions of regular and evolved stars for Oosterhoff type I (OoI) and type II (OoII) clusters. The proximity of the NGC~6401 RRab and RRc locations to the lines corresponding to OoI clusters confirms our classification. We do not present the corresponding SDSS-$g\arcmin$ band diagram since we have no comparison theoretical lines to display.

\subsection{Correcting for extinction}
\label{sec:reddening} 
As pointed out in \citet{Barbuy99}, NGC 6401 suffers from heavy differential reddening. Foreground reddening estimates at and around the cluster centre coordinates range from $E(B-V)=0.53$ \citep{Barbuy99} to $E(B-V)=1.1$ \citep{Valenti07}, with multiple authors giving estimates in between those extremes (e.g. $E(B-V)=0.85$ \citep{Harris96, Piotto02}, 0.81 \citep{Bica86, Schlafly11}, 0.59 \citep{Minniti95}, 0.96 \citep{Schlegel98}). These can be converted to $E(V-I)$ reddening values through $E(V-I)=1.259 E(B-V)$, derived from \citet{Schlegel98}.
In order to derive reasonable distances to the cluster, appropriate reddening calibrations need to be applied. 

After taking into account small corrections for period and metallicity differences between stars, \citet{Sturch66} discovered that all RRab stars have nearly the same dereddened $(B-V)$ colour at minimum light (i.e. between the phase interval 0.5-0.8). \citet{Mateo95} suggested that $(V-I)$ colour at minimum light is a better indicator of foreground reddening. This was confirmed by \citet{Day02}, who found that the mean dereddened $(V-I)$ colour at minimum light was $0.57\pm 0.025$ mag, with little-to-no dependence on metallicity or period. This was further refined by \citet{Guldenschuh05} to $\overline{(V-I)_{0,{\textrm{min}}}}=0.58\pm 0.02$ mag, who also found no dependence on metallicity or pulsation amplitude. Guldenschuh also provided an independent colour-metallicity calibration in the form of $\overline{(V-I)_{0,{\textrm{min}}}}=0.569-0.008$[Fe/H], with an rms of 0.024 mag. \citet{Kunder10} validated these relationships for RRab stars in the Galactic bulge.

To determine distances to individual RRab stars, we calculated the mean dereddened magnitudes using $I_0=I-A_I$ and $V_0=V-A_V$.
The value for the extinction $A_I$ was obtained using the relationships in 
\citet{Schlegel98}:
\begin{equation}
A_I = 1.4626E(V-I),
\end{equation}
and, again following \citet{Schlegel98}, $A_V=1.6835A_I$.
The reddening in the optical is given by
\begin{equation}
E(V-I)=(V-I)-(V-I)_0,
\end{equation} 
where the color $(V-I)$ is evaluated from the observed RRab colour at minimum light, and $(V-I)_0$ is the intrinsic colour at minimum light $\overline{(V-I)_{0,{\textrm{min}}}}=0.58$ mag. For comparison, we also provide the reddening values obtained using the following approximate formula from \citet{Kovacs01}:
\begin{equation}
\label{KoW01}
(V-I)_0= 0.253 \mbox{log}(P) - 0.388 A_1 + 0.364 A_3 + 0.648.
\end{equation}
The reddening values evaluated using the minimum light method are given in column 6 of Table~\ref{tab:phys_rrab} and those calculated via expression~\ref{KoW01} in column 7.
\begingroup
\setlength{\tabcolsep}{2pt} 
\begin{table}
\scriptsize
\begin{center}
\caption[] {\small Physical parameters for the RRab stars in NGC 6401. Stars marked with an asterisk (*) were not included when estimating averages. Column 6 lists the reddening calculated using the minimum light method, while column 7 lists the values obtained using equation~~\ref{KoW01}.}
\hspace{0.001cm}
 \begin{tabular}{lcccccccl}
\hline 
Star & [Fe/H]$_{\mathrm{ZW}}$ & $M_{\mathrm{V}}$ & log$(L/{L_{\odot}})$ & $T_{\rm eff}$ & $E(V-I)$ & $E(V-I)$ & Distance\\
     &               &       &                          &               &        & [KW01] &  (kpc) \\
\hline
 &  & & RRab stars & & & &\\
\hline
V2   & -1.212 & 0.450 & 1.725 & 6344.22 & 1.292 & 1.339   & 6.406\\ 
V5   & -1.287 & 0.638 & 1.651 & 6522.05 & 1.405 & 1.491   & 6.720\\ 
V6   & -1.305 & 0.708 & 1.624 & 6654.18 & 1.176 & 1.305   & 7.419\\ 
V7   & -1.258 & 0.639 & 1.651 & 6504.83 & 1.146 & 1.224   & 6.726\\ 
V9   & -1.419 & 0.709 & 1.627 & 6502.29 & 1.344 & 1.422   & 4.993\\ 
V10  & -1.248 & 0.637 & 1.651 & 6552.30 & 1.249 & 1.340   & 6.808\\ 
V11  & -1.196 & 0.634 & 1.651 & 6429.07 & 1.114 & 1.171   & 6.970\\ 
V13  & -1.140 & 0.623 & 1.654 & 6498.34 & 1.257 & 1.327   & 7.306 \\ 
V14* & -1.444 & 0.652 & 1.650 & 6250.92 & 1.248 & 1.286   & 5.899\\ 
V15  & -1.270 & 0.676 & 1.636 & 6564.51 & 1.210 & 1.302   & 6.866\\ 
V16  & -1.210 & 0.692 & 1.628 & 6565.09 & 1.424 & 1.512   & 7.278 \\ 
V17  & -1.325 & 0.659 & 1.644 & 6561.62 & 1.177 & 1.276   & 5.396\\ 
V18  & -1.315 & 0.592 & 1.671 & 6460.35 & 1.218 & 1.294   & 6.670\\ 
V19  & -1.213 & 0.636 & 1.651 & 6480.19 & 1.126 & 1.196   & 6.972\\ 
V24  & -1.303 & 0.701 & 1.627 & 6517.09 & 1.183 & 1.259   & 6.407\\ 
V25  & -1.277 & 0.722 & 1.618 & 6415.69 & 1.486 & 1.531   & 4.302\\ 
V26  & -1.227 & 0.676 & 1.635 & 6599.57 & 1.469 & 1.570   & 6.088\\ 
V28* & -1.350 & 0.631 & 1.656 & 6295.70 & 1.249 & 1.284   & 7.219\\ 
V31  & -1.210 & 0.631 & 1.653 & 6438.13 & 1.487 & 1.548   & 5.516\\ 
V32  & -1.135 & 0.554 & 1.682 & 6375.44 & 1.508 & 1.546   & 5.430\\ 
V34  & -1.277 & 0.704 & 1.625 & 6469.60 & 1.204 & 1.263   & 6.121\\ 
V37  & -1.247 & 0.684 & 1.633 & 6622.74 & 1.269 & 1.379   & 6.667\\ 
\hline
Weighted &  -1.254     & 0.648       & 1.647       & 6503.87     & 1.287       & 1.365	     & 6.353   \\
Mean     & $\pm$ 0.064 & $\pm$ 0.062 & $\pm$ 0.024 & $\pm$ 78.37 & $\pm$ 0.128 & $\pm$ 0.124  & $\pm$ 0.814 \\
\hline
\hline
\label{tab:phys_rrab}
\end{tabular}
\end{center}
\end{table}
\endgroup

\begin{table}
\scriptsize
\begin{center}
\caption[] {\small Physical parameters for the RRc stars in NGC 6401. Stars marked with an asterisk (*) were not included when estimating averages.}
\hspace{0.001cm}
 \begin{tabular}{lccccc}
\hline 
Star & [Fe/H]$_{\mathrm{ZW}}$ & $M_{\mathrm{V}}$ & log$(L/{\rm L_{\odot}})$ & $T_{\rm eff}$  \\
\hline
 & & RRc stars & & \\
\hline
 V8   & -1.086 & 0.566 & 1.676 & 7398.51 \\ 
 V12* & -1.408 & 0.584 & 1.676 & 7403.00 \\ 
 V21  & -1.269 & 0.566 & 1.680 & 7447.49 \\ 
 V22  & -1.327 & 0.600 & 1.668 & 7402.74 \\ 
 V23  & -1.489 & 0.514 & 1.706 & 7282.88 \\ 
 V27  & -1.043 & 0.588 & 1.666 & 7464.05 \\ 
 V29  & -1.282 & 0.598 & 1.668 & 7427.72 \\ 
 V30* &  0.802 & 0.663 & 1.592 & 7707.52 \\ 
 V33  & -0.911 & 0.536 & 1.683 & 7396.88 \\ 
 V35  & -1.635 & 0.626 & 1.665 & 7370.81 \\ 
 V36  & -1.226 & 0.587 & 1.671 & 7388.00 \\ 
\hline
Weighted & -1.252  & 0.576 & 1.676 & 7397.68 \\
Mean & $\pm$ 0.210 & $\pm$ 0.032 & $\pm$ 0.012 & $\pm$ 49.23\\
\hline
\hline
\label{tab:phys_rrc}
\end{tabular}
\end{center}
\end{table}

\subsection{Distance to NGC6401 using RR Lyrae stars}
The absolute magnitude $M_{\mathrm{V}}$ can be used to determine the distance $d$ to each RRab star using:
\begin{equation}
d = 10^{1+0.2(V_0-M_{\mathrm{V}})},
\end{equation}
\noindent where $V_0$ is the star's dereddened magnitude.
We obtain a mean distance to the cluster $d\approx 6.35 \pm 0.81$ kpc. The derived distance is consistent with the value reported in \citet{Valenti07} ($d\approx 7.7$ kpc) and significantly lower than the one found by \citet{Barbuy99} ($d\approx 12.0 \pm 1.0$ kpc). Our distances were estimated using the intrinsic colour at minimum light method.

\section{Discussion}
\label{sec:discussion}
This work presents the first detailed photometric time-series study of NGC~6401 employing CCD data. 
NGC~6401 is a little-investigated globular cluster $5.3\degr$ from the Galactic centre. It is affected by strong differential reddening, which complicates photometric calibration, and hence the estimation of the physical parameters. This is possibly one of the reasons it has not been extensively studied in the past.

Using the RR Lyrae stars that are the most plausible cluster members, we have determined that NGC~6401 is an OoI cluster of intermediate metallicity that belongs to the metal-rich population of $\sim$20 bulge clusters that lie within 2 kpc from the Galactic centre \citep{Minniti95b}. \citet{Bica16} conclude that a cut-off of 3 kpc from the Galactic Centre is suitable in terms of isolating a bona-fide bulge cluster sample, with little contamination. As pointed out by \citet{Rossi15}, inner bulge globular clusters appear to be trapped in the bulge bar due to its high mass. This is very likely the case for NGC~6401 as well, since it is also in the bulge \citep{Zoccali16}.

\citet{Forbes01} argue that the mean metallicity of globular clusters reveals information about the mass of the host galaxy because the chemical enrichment of globular clusters is directly linked to the galaxy formation process. In particular, they claim that the inner metal-rich globular clusters and the bulge field stars have similar colours because they formed at the same time from the same chemically enriched gas. Their findings agree with those of \citet{Barbuy98}, who also conclude that bulge clusters share comparable properties with the bulge field stars, including not only metal-rich globular clusters, but also intermediate metallicity ones. In this work, we have derived independent metallicity and distance estimates for NGC~6401, which are compatible with the characteristics of other metal-rich globular clusters in the Galactic bulge. Our results provide further support to the idea that the bulge field stars and the inner globular cluster population evolved on the same timescales. 


\section{Summary}
\label{sec:summary}
We have updated the census of variable stars in NGC 6401. We use three different methods to recover 34 RR Lyrae cluster variables, three of which do not appear in the OGLE database (V8, V10 and V14). We do not find evidence of variability for the previously reported variables V1,  V20 and OGLE-BLG-RRLYR-23724. We discovered 13 new variables, E1 and LPV1-12 and classified V3 as a W Virginis star for the first time. We detected double-mode non-radial pulsations in the RRc star V12, adding another example of these rare objects in globular clusters. Finally, we used the RR Lyrae content to establish that NGC6401 is an OoI type cluster.

The parameters obtained through Fourier decomposition of the light curves of a selected subset of RRab and RRc stars were used to estimate the metallicity, absolute magnitude, luminosity and effective temperature of each variable. The average metallicity for the cluster derived using the RRab stars ${\rm [Fe/H]}_{\mathrm{ZW}} = -1.25 \pm 0.06$ (${\rm [Fe/H]}_{\mathrm{UVES}} = -1.13 \pm 0.06$) is consistent with that evaluated from the RRc stars ${\rm [Fe/H]}_{\mathrm{ZW}} = -1.25 \pm 0.21$ and lower than the estimate by \citet{Barbuy99}. 

The mean distance to the cluster, estimated after individual RRab stars were corrected for differential reddening, is $d\approx 6.35 \pm 0.81$ kpc.

\section*{Acknowledgments}
We thank Igor Soszynski for providing calibrated OGLE light curves of stars in our field. 
This work made use of Python routines based on gatspy and astroML methods.
This project was supported by DGAPA-UNAM grant through project IN106615. 
D.M.B. acknowledges NPRP grant \# X-019-1-006 from the Qatar National Research Fund (a member of Qatar Foundation).
This work makes use of observations from the LCOGT network.
This work has made a large use of the SIMBAD and ADS services.

\noindent
----------------------------------------------------------------------------
\noindent
\\$^{1}$Astronomisches Rechen-Institut, Zentrum f{\"u}r Astronomie der Universit{\"a}t Heidelberg (ZAH), 69120 Heidelberg, Germany.
\\$^{2}$Instituto de Astronom\1a, Universidad Nacional Aut\'onoma de M\'exico.
Ciudad Universitaria CP 04510, Mexico.
\\$^{3}$Qatar Environment and Energy Research Institute(QEERI), HBKU, Qatar Foundation, Doha, Qatar.
\\$^{4}$European Southern Observatory, Karl-Schwarzschild-Stra$\beta$e 2, 85748 Garching bei M\"{u}nchen, Germany.
\\$^{5}$SUPA, School of Physics and Astronomy, University of St. Andrews, North Haugh, St Andrews, KY16 9SS, United Kingdom.
\\$^{6}$Space Telescope Institute, 3700 San Martin Drive, Baltimore, MD 21218, USA.
\\$^{7}$Las Cumbres Observatory Global Telescope Network, 6740 Cortona Drive, suite 102, Goleta, CA 93117, USA.
\\$^{8}$Niels Bohr Institute \& Centre for Star and Planet Formation, University of Copenhagen, {\O}ster Voldgade 5, 1350 - Copenhagen K, Denmark.
\\$^{9}$Planetary and Space Sciences, Department of Physical Sciences, The Open University, Milton Keynes, MK7 6AA, UK.
\\$^{10}$Max Planck Institute for Solar System Research, Max-Planck-Str. 2, 37191 Katlenburg-Lindau, Germany.

\newpage
\appendix
\section{Supplementary light curves}
\begin{figure*}
  \begin{tabular}{@{}cccc@{}}
    \includegraphics[width=.22\textwidth]{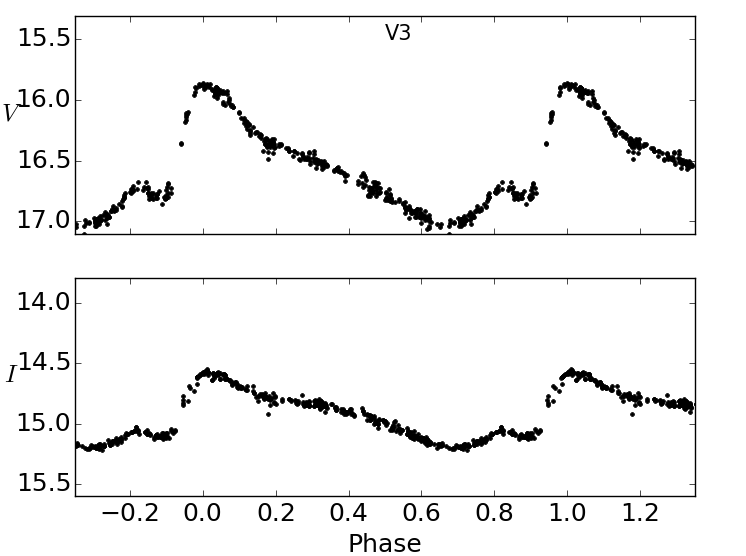}		 &
    \includegraphics[width=.23\textwidth]{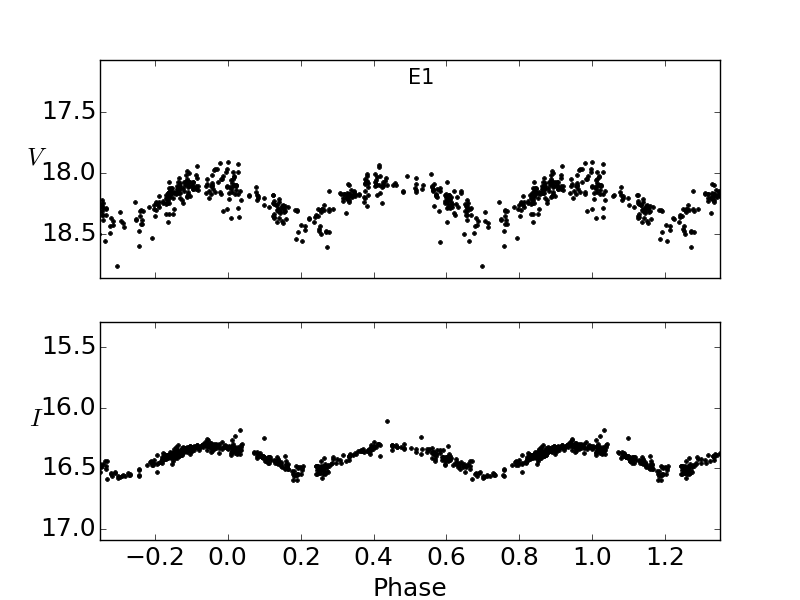}		 &
    \includegraphics[width=.23\textwidth]{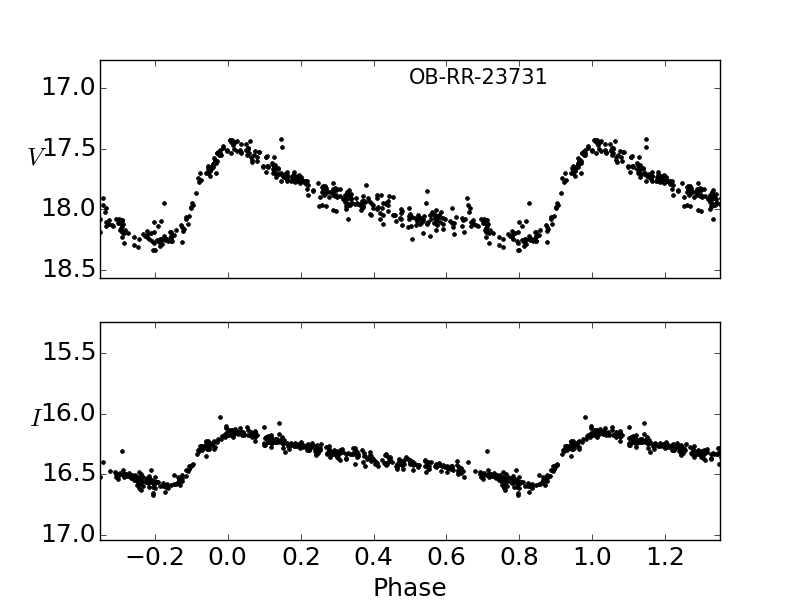}	 &
    \includegraphics[width=.23\textwidth]{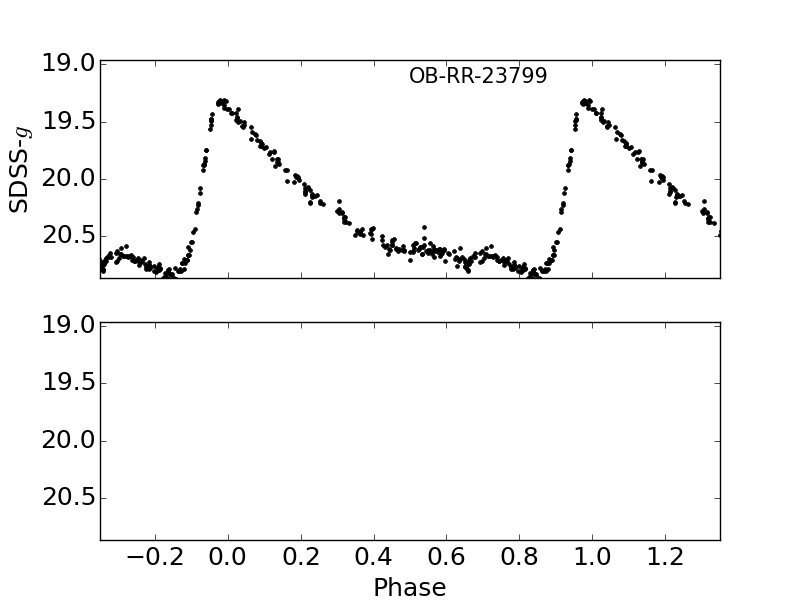}	 \\
    \includegraphics[width=.23\textwidth]{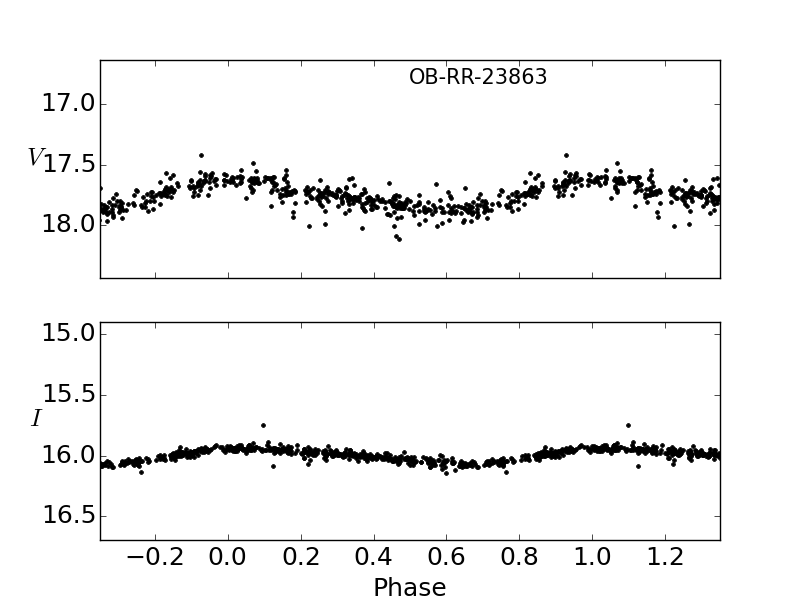}	 &
    \includegraphics[width=.23\textwidth]{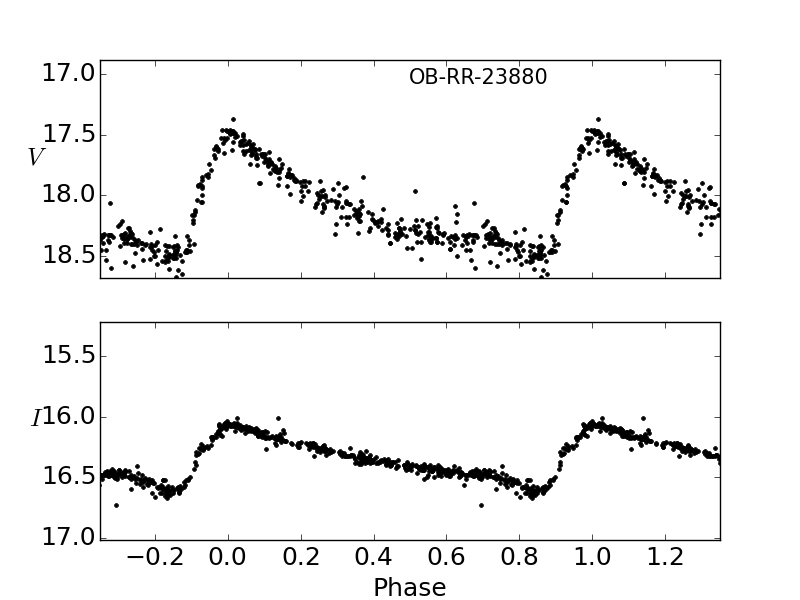}	 &
    \includegraphics[width=.23\textwidth]{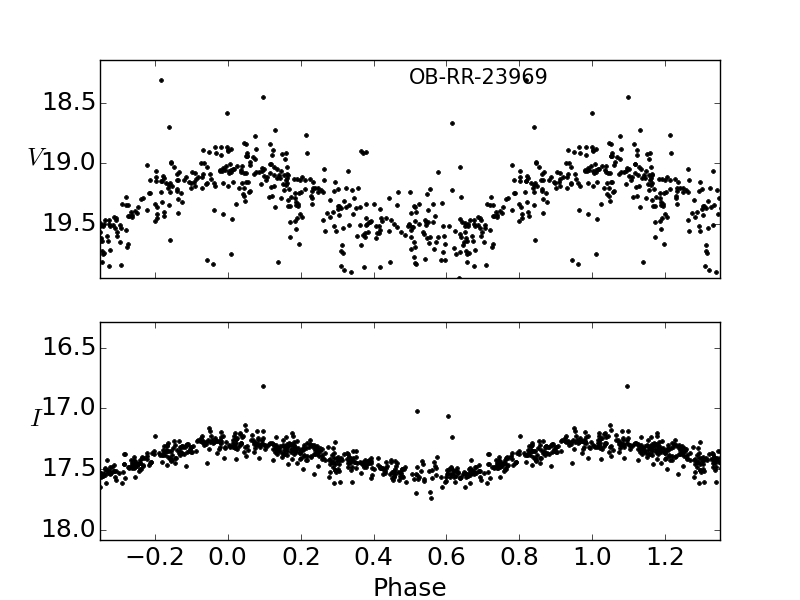}	 &
    \includegraphics[width=.23\textwidth]{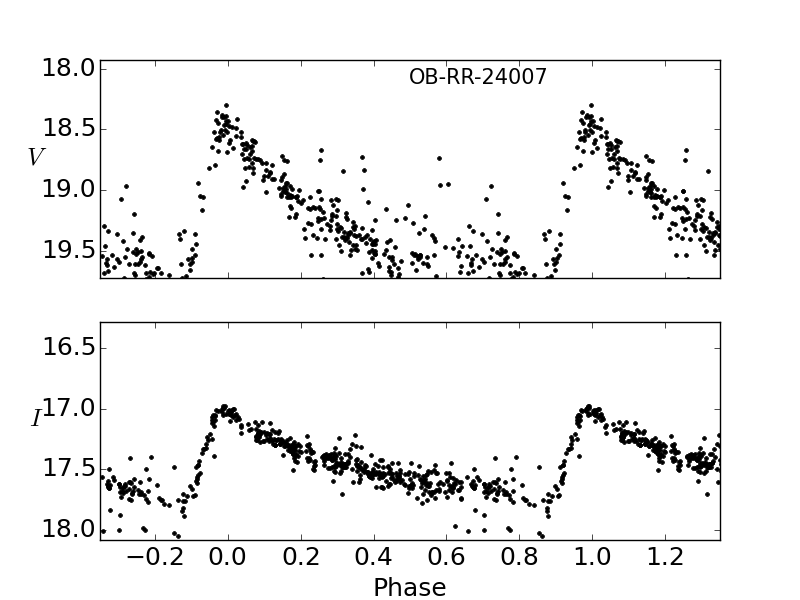}	 \\
    \includegraphics[width=.23\textwidth]{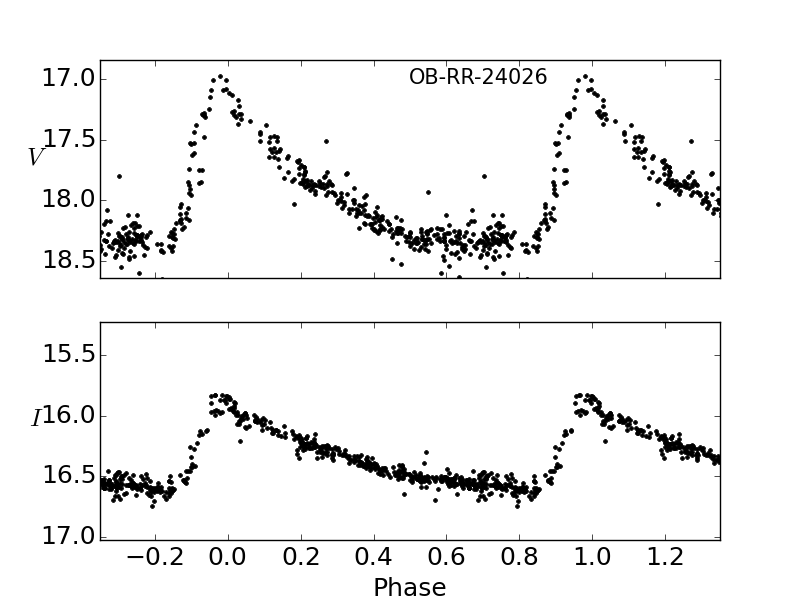}	 &
    \includegraphics[width=.23\textwidth]{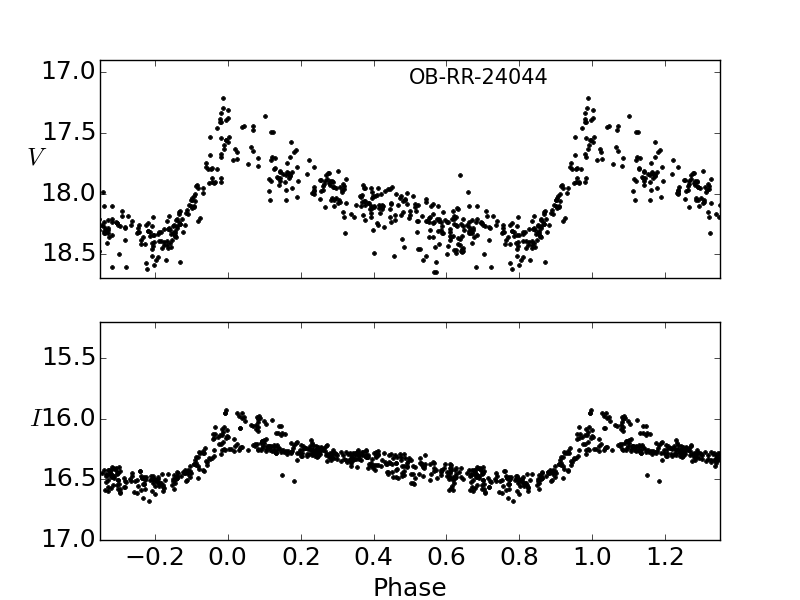}	 &
    \includegraphics[width=.23\textwidth]{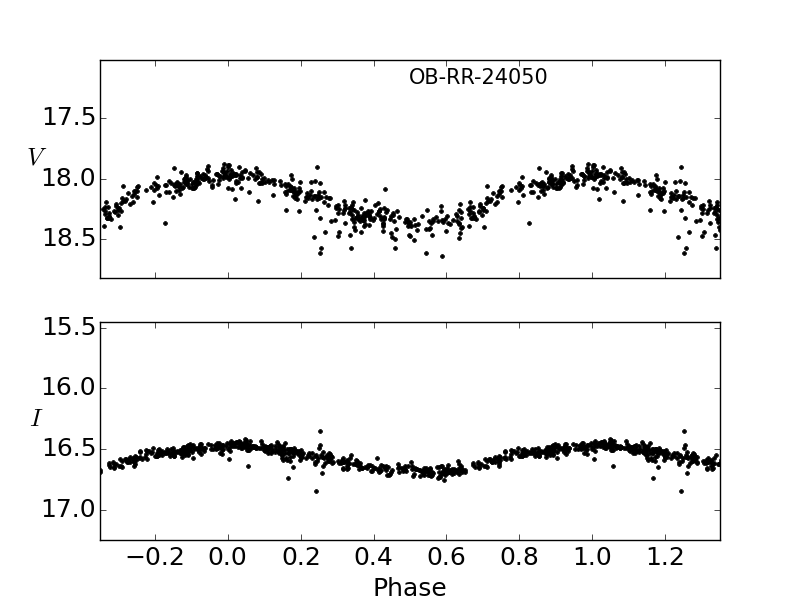}	 &
    \includegraphics[width=.23\textwidth]{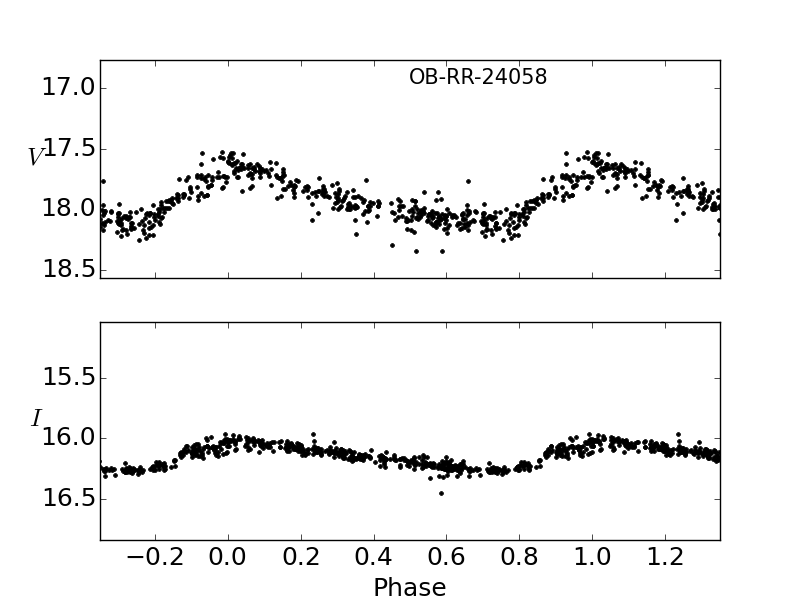}	 \\
    \includegraphics[width=.23\textwidth]{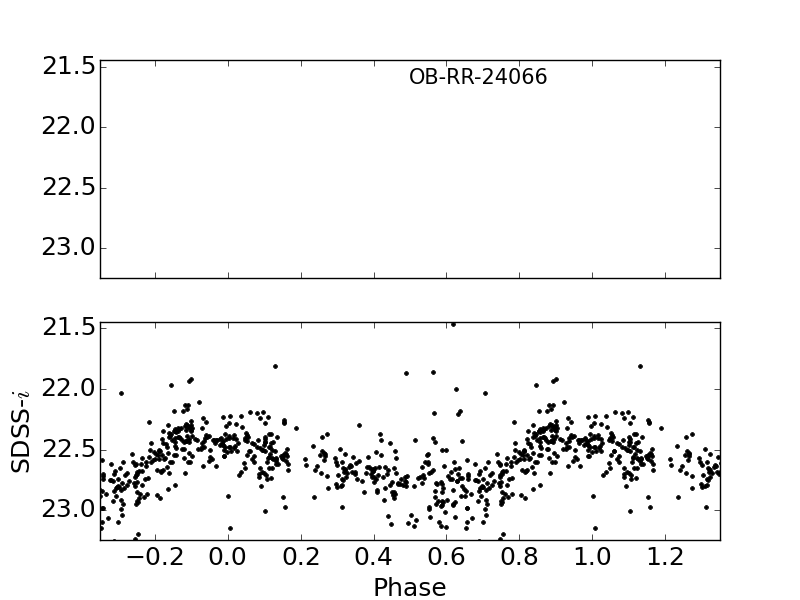}	 &
    \includegraphics[width=.23\textwidth]{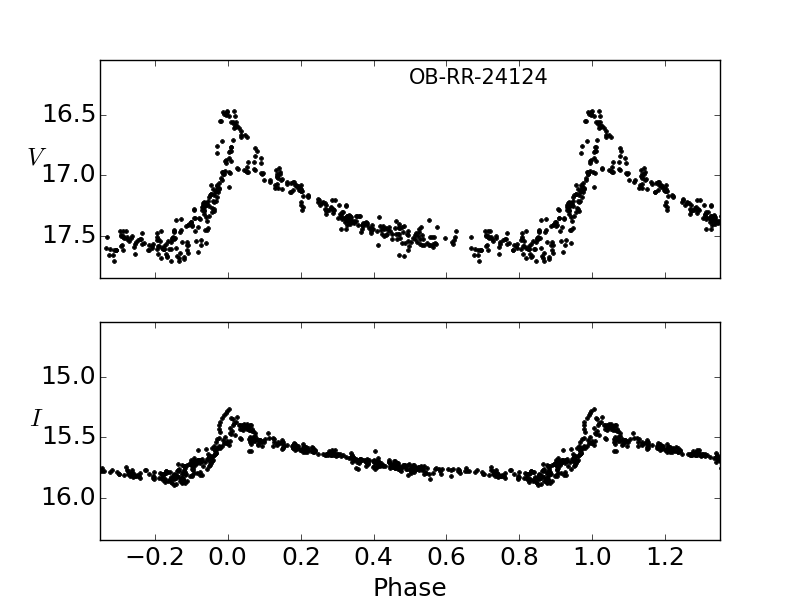}	 &
    \includegraphics[width=.23\textwidth]{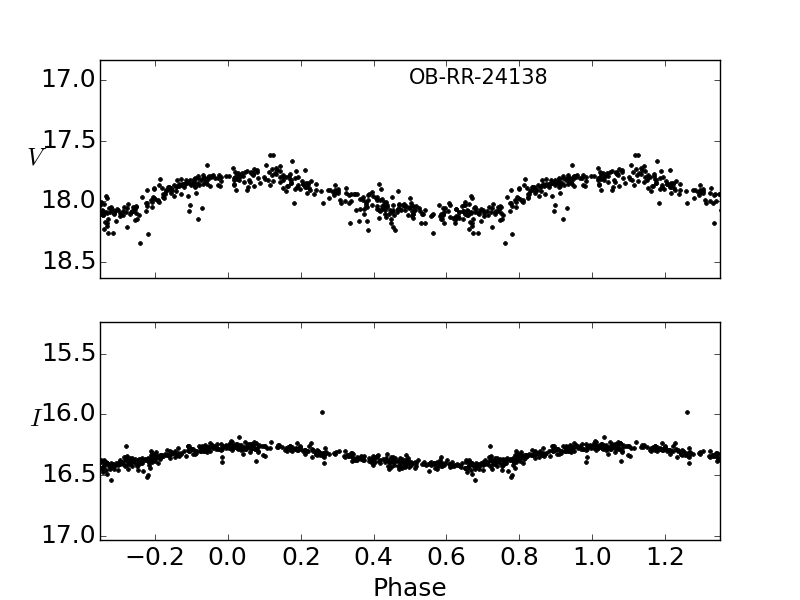}	 &
    \includegraphics[width=.23\textwidth]{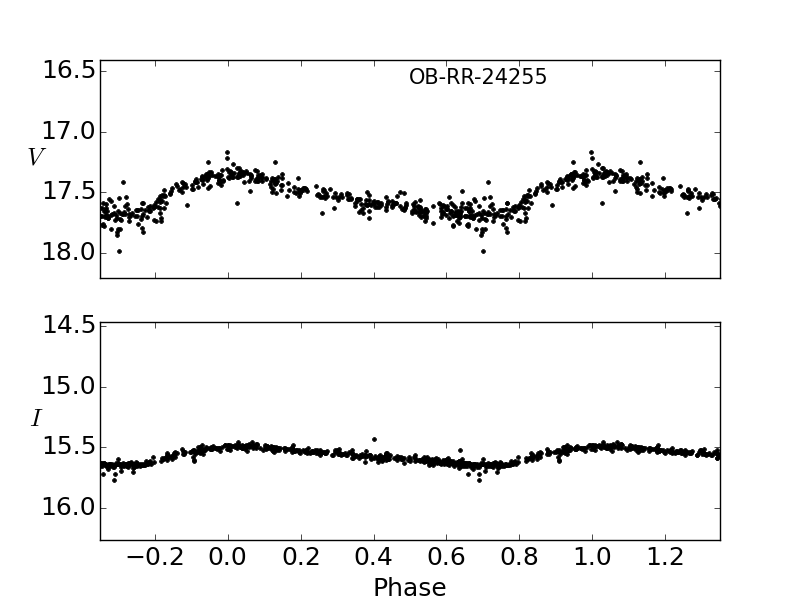}	 \\
    \includegraphics[width=.23\textwidth]{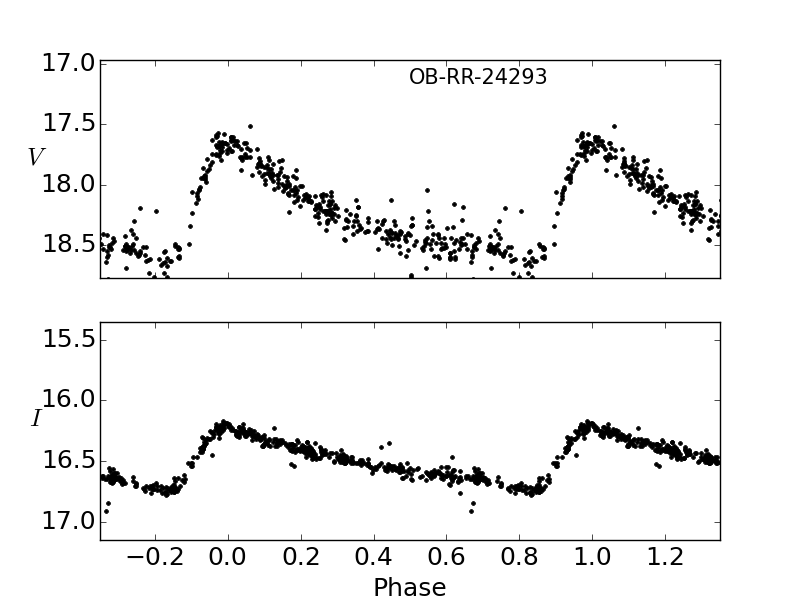}	 &
    \includegraphics[width=.23\textwidth]{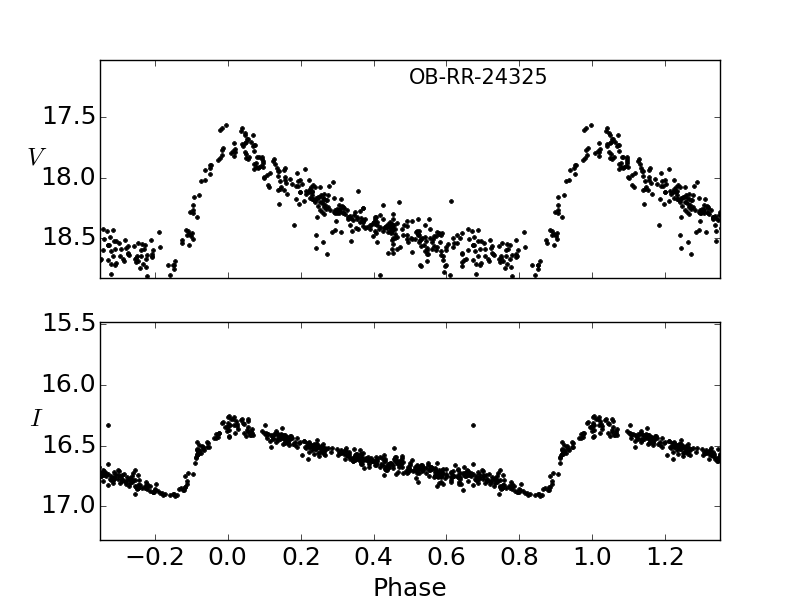}	 &
    \includegraphics[width=.23\textwidth]{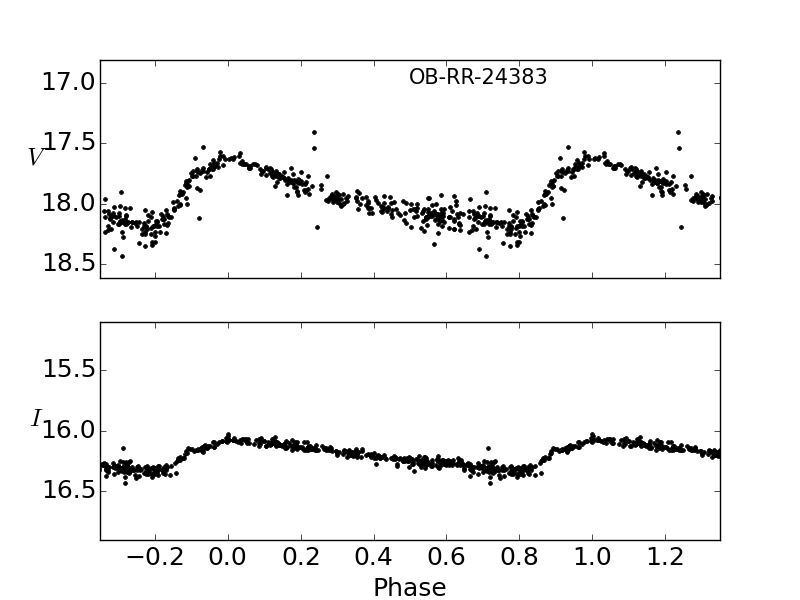}	 &
  \end{tabular}
  \caption{Standard SDSS-$g\arcmin$ and SDSS-$i\arcmin$ band light curves, linearly transformed to the $V$ and $I$ bands respectively, for V3, E1 and the RR Lyrae variable stars listed in Table~\ref{tab:var_out}. Note that no SDSS-$i\arcmin$ band data were available for OB-RR-23799 so only SDSS-$g\arcmin$ band instrumental magnitudes are displayed, and vice versa for OB-RR-24066.}
  \label{fig:varout1}
\end{figure*}

\begin{figure*}
  \begin{tabular}{@{}cccc@{}}
    \includegraphics[width=.23\textwidth]{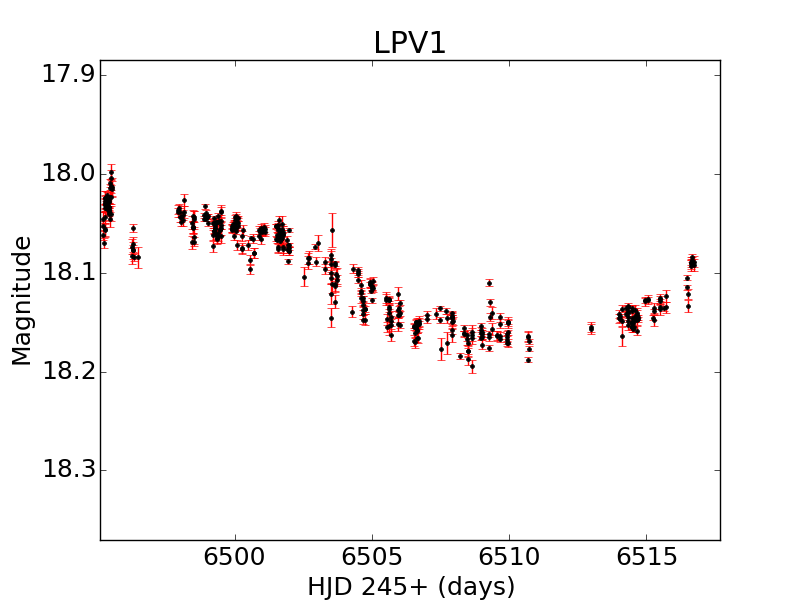}  	      &
    \includegraphics[width=.23\textwidth]{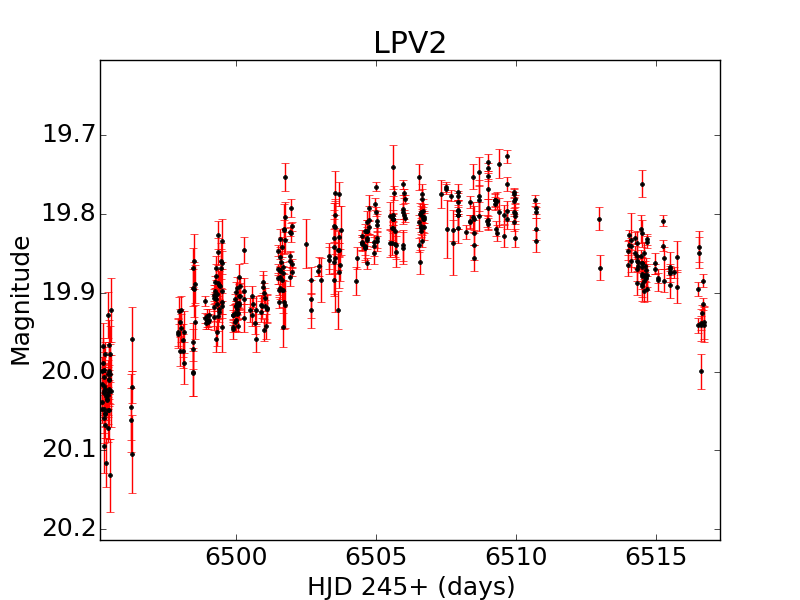}  	      &
    \includegraphics[width=.23\textwidth]{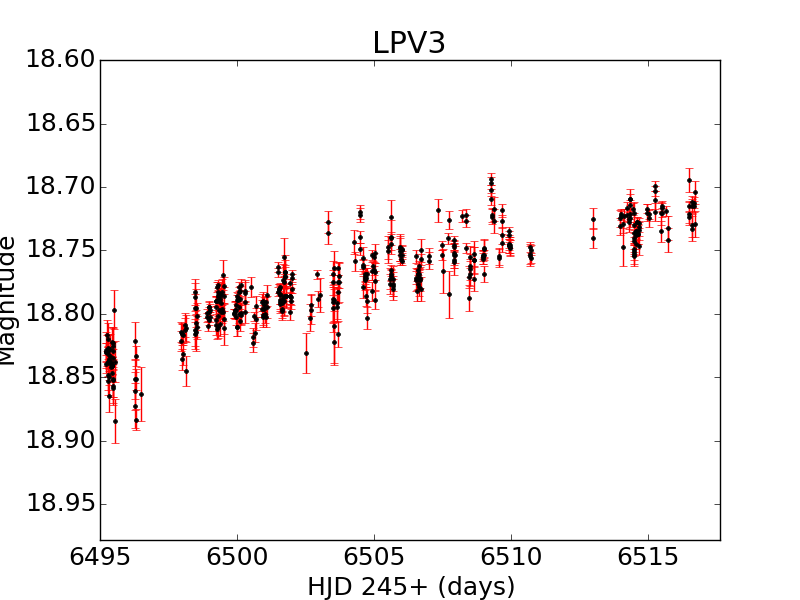}  	      &
    \includegraphics[width=.23\textwidth]{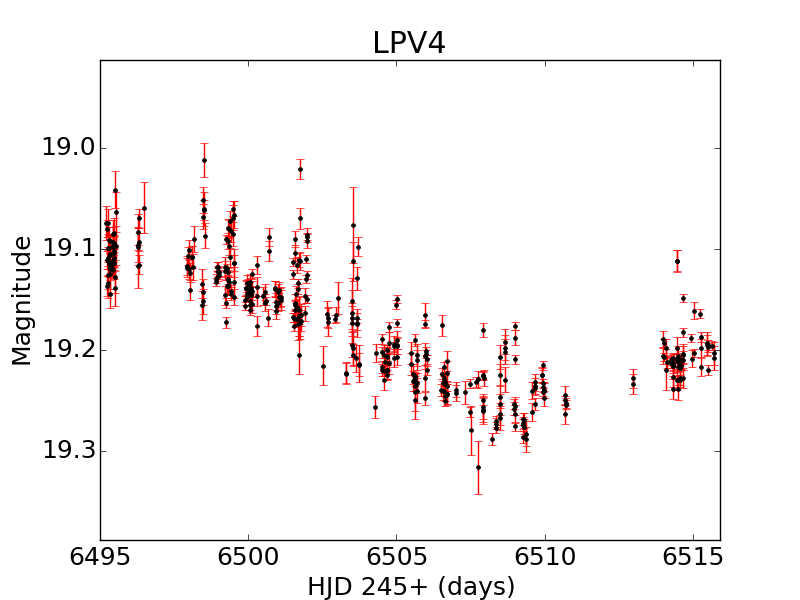}  	      \\
    \includegraphics[width=.23\textwidth]{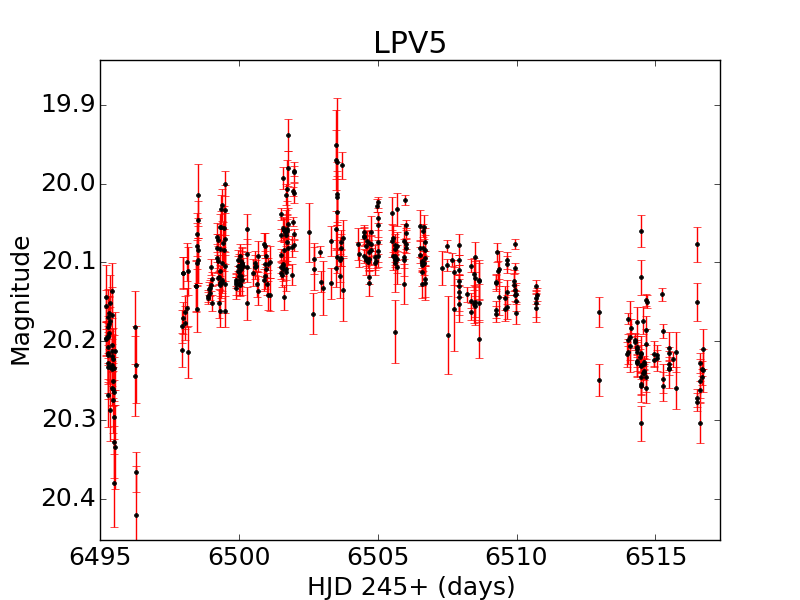}  	      &
    \includegraphics[width=.23\textwidth]{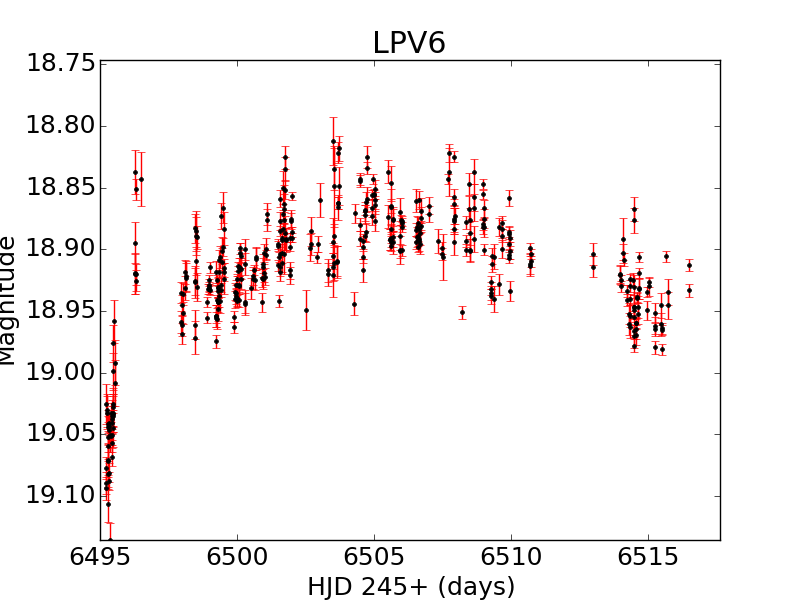}  	      &
    \includegraphics[width=.23\textwidth]{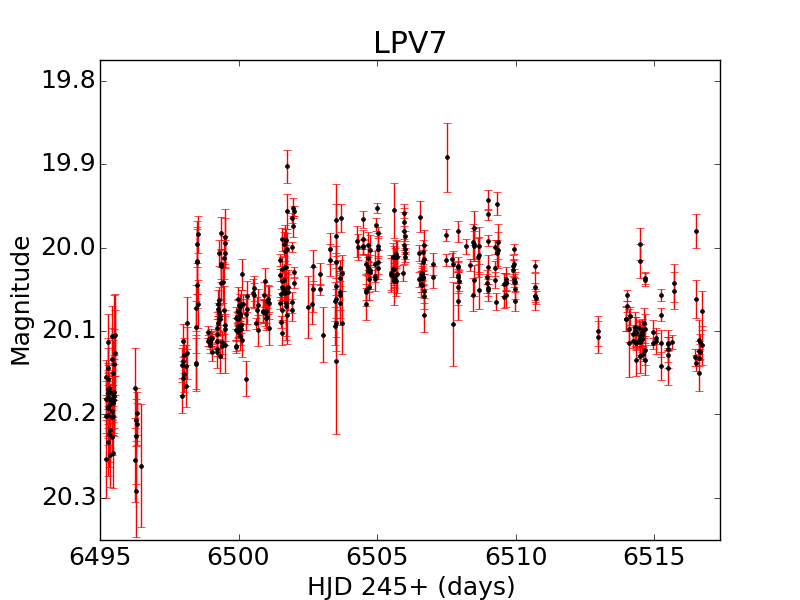}  	      &
    \includegraphics[width=.23\textwidth]{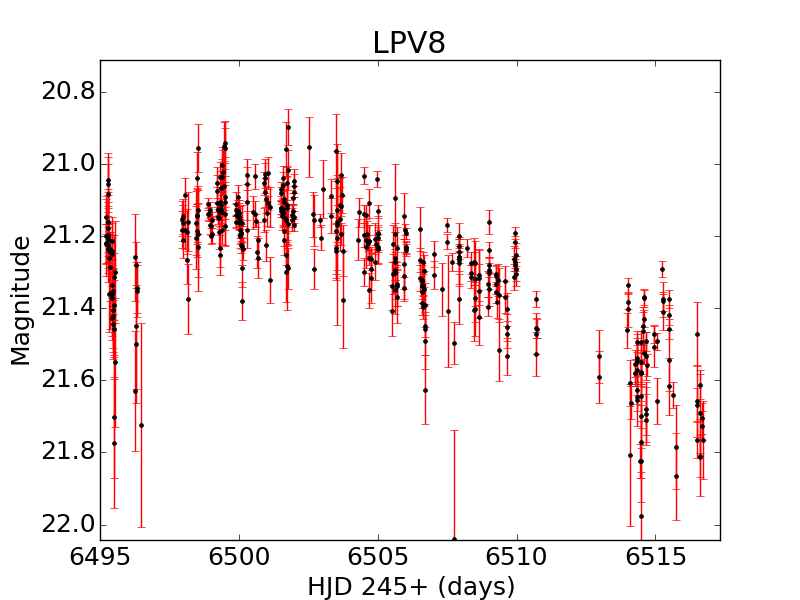}  	      \\
    \includegraphics[width=.23\textwidth]{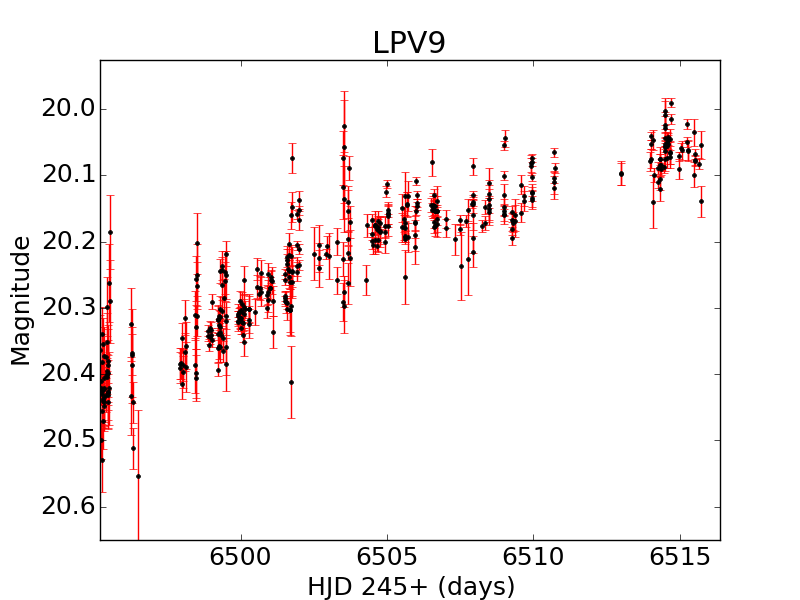}  	      &
    \includegraphics[width=.23\textwidth]{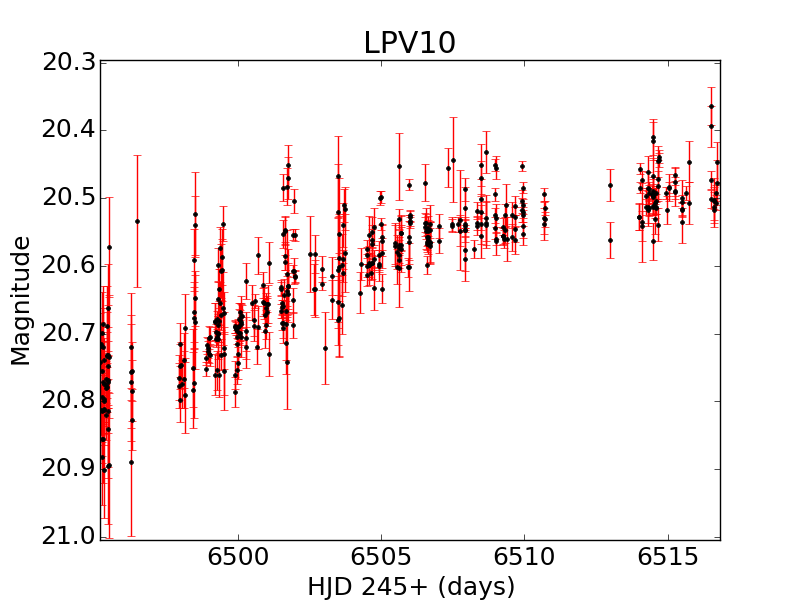} 	      &
    \includegraphics[width=.23\textwidth]{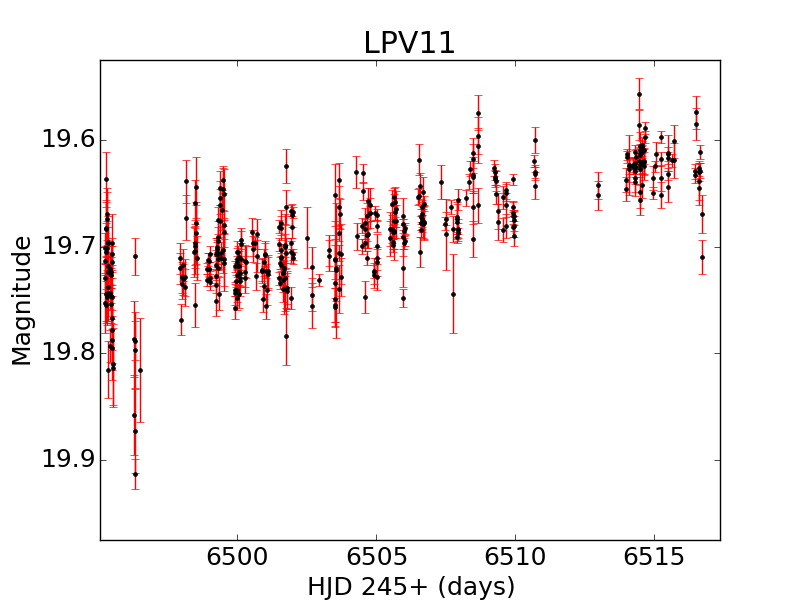} 	      &
    \includegraphics[width=.23\textwidth]{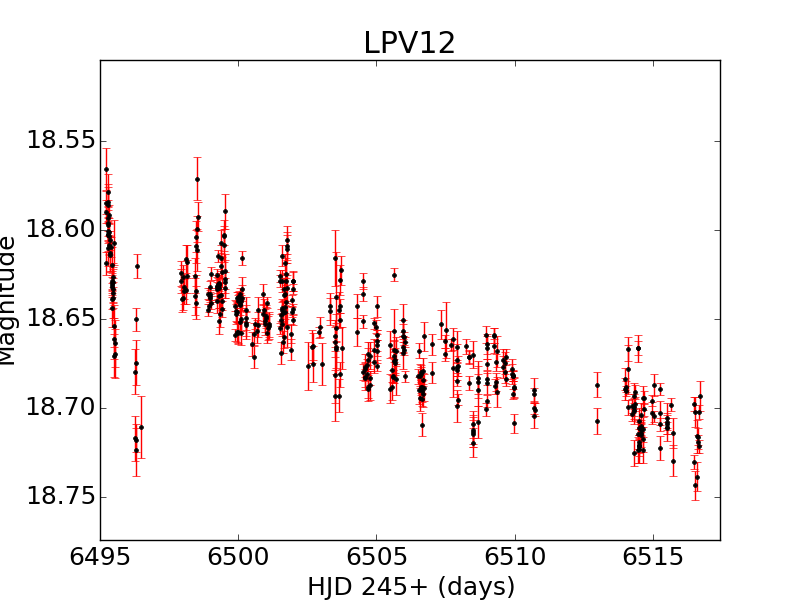} 	      \\
    \includegraphics[width=.23\textwidth]{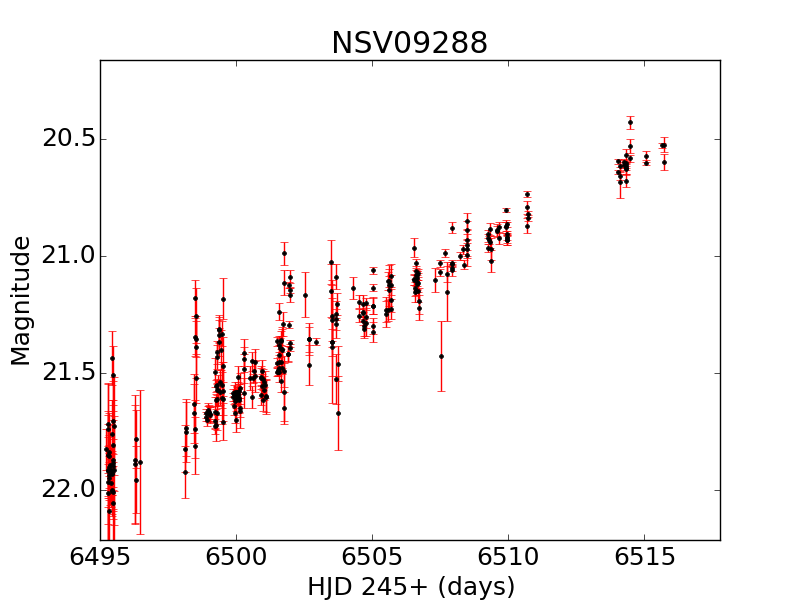}	       &
    \includegraphics[width=.23\textwidth]{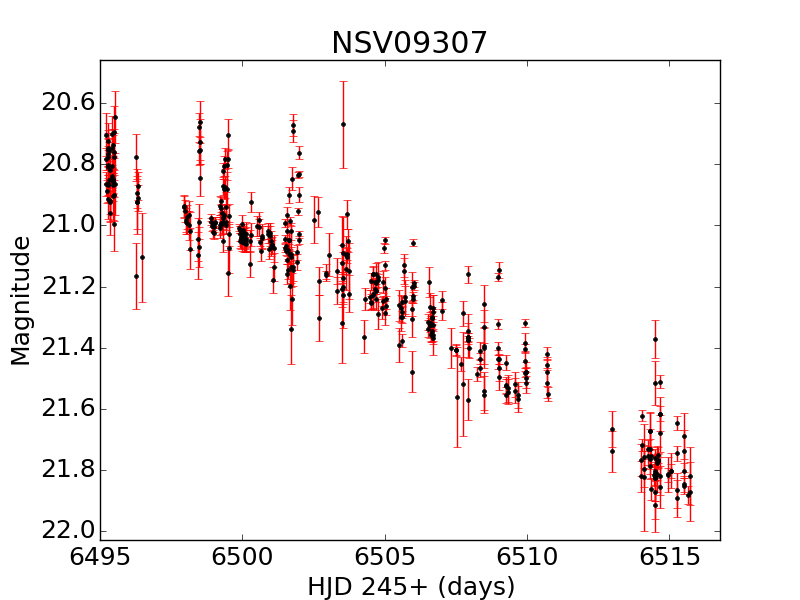}	       &
    \includegraphics[width=.23\textwidth]{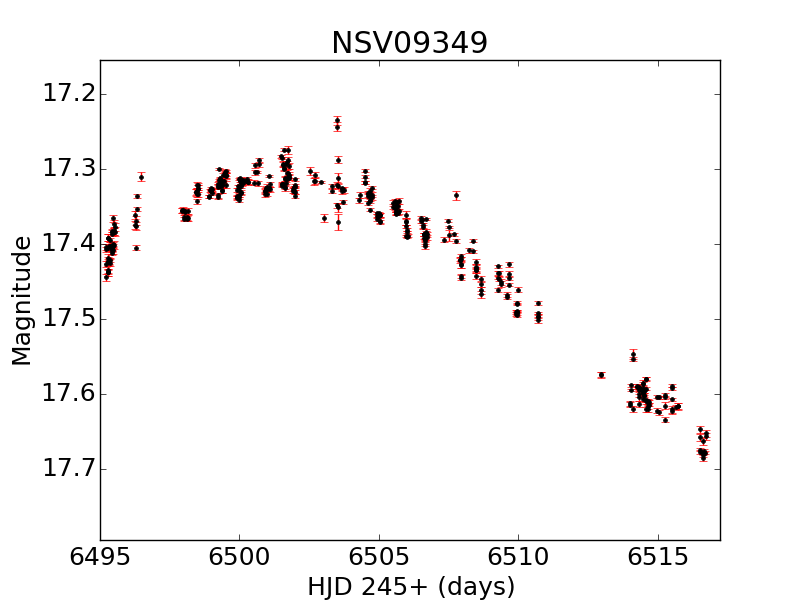}	       &
    \includegraphics[width=.23\textwidth]{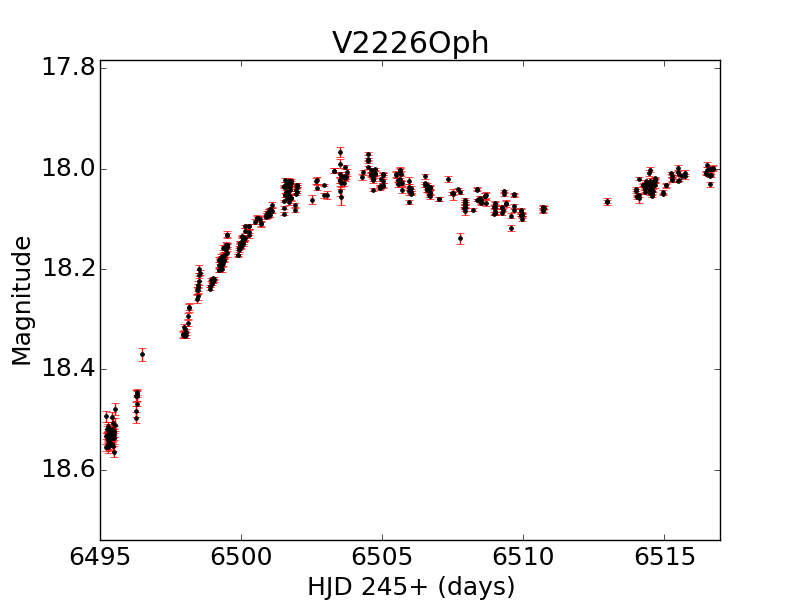}	       \\
    \includegraphics[width=.23\textwidth]{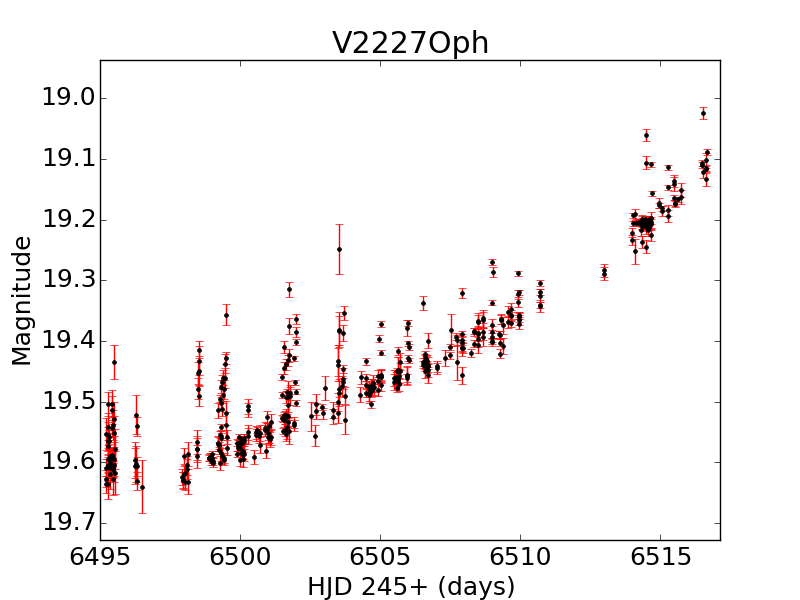}	       &
  \end{tabular}
  \caption{Light curves of the long period variable stars listed in Table~\ref{tab:var_out}. The reported magnitudes are in the instrumental SDSS-$g\arcmin$ scale.}
  \label{fig:varout2}
\end{figure*}

\end{document}